\pdfoutput=1
\documentclass[10pt]{jpaper}

\usepackage{etex}
\usepackage[sort,nocompress,adjust]{cite} 

\usepackage{amsmath,amssymb,amsfonts}
\usepackage{algorithmic}
\usepackage{graphicx}
\usepackage{listings}
\usepackage{xcolor}
\definecolor{olive}{rgb}{0.33, 0.42, 0.18}
\usepackage{makecell}
\usepackage{textcomp}
\usepackage{fancyhdr}
\usepackage{tabu}
\usepackage{enumitem}
\usepackage[normalem]{ulem}
\usepackage{inconsolata}
\usepackage[scaled]{beramono}
\usepackage{float}
\usepackage{caption}
\definecolor{keywordcolor}{HTML}{cc33ff}

\def\BibTeX{{\rm B\kern-.05em{\sc i\kern-.025em b}\kern-.08em
T\kern-.1667em\lower.7ex\hbox{E}\kern-.125emX}}

\usepackage{verbatim}   

\usepackage{wrapfig}
\usepackage{multirow}
\usepackage{balance}
\usepackage{cite}

\usepackage{pifont}

\newcommand{\smallcapital}{\fontsize{9pt}{10pt}\selectfont}

\usepackage{rotating}

\usepackage{arydshln}

\usepackage{placeins}

\usepackage{gensymb}

\usepackage{enumitem}
\usepackage[ruled,vlined]{algorithm2e}
\usepackage{upgreek,textgreek}
\usepackage[htt]{hyphenat}
\usepackage{appendix}

\hypersetup{colorlinks,
	linkcolor=blue,
	citecolor=blue,
	urlcolor=blue,
	filecolor=blue}

\usepackage{mathtools}

\usepackage{tikz}

\usepackage{listings}
\usepackage[utf8]{inputenc}
\lstdefinestyle{customcpp}{
	 aboveskip=0in,
	  belowskip=0in,
	   abovecaptionskip=0in,
	    belowcaptionskip=0in,
	     captionpos=b,
	      xleftmargin=\parindent,
	       language=C++,
	        morekeywords={forall},
		 showstringspaces=false,
		  basicstyle={\linespread{0.6}\fontseries{sb}\small\ttfamily},
		   keywordstyle=\bfseries,
		    commentstyle=\itshape\color{green!40!black},
	    }

\begin{document}

%

\title{A Hardware-Software Stack for Serverless Edge Swarms}

\author{Liam Patterson, David Pigorovsky, Brian Dempsey, Nikita Lazarev,\\ Aditya Shah, Clara Steinhoff, Ariana Bruno, Justin Hu, \\and Christina Delimitrou\\Cornell University}

%

\date{}

\maketitle

\pagestyle{plain}

\begin{abstract}
	{Swarms of autonomous devices are increasing in ubiquity and size, making the need for rethinking their hardware-software system stack critical. 


We present HiveMind, the first swarm coordination platform that enables programmable execution of complex task workflows between cloud and edge resources in a performant and scalable manner. HiveMind is a software-hardware platform that includes a domain-specific language to simplify programmability of cloud-edge applications, a program synthesis tool to automatically explore task placement strategies, a centralized controller that leverages serverless computing to elastically scale cloud resources, and a reconfigurable hardware acceleration fabric for network and remote memory accesses. 

We design and build the full end-to-end HiveMind system on two real edge swarms comprised of drones and robotic cars. 
We quantify the opportunities and challenges serverless introduces to edge applications, as well as the trade-offs between centralized and distributed coordination. We show that HiveMind achieves significantly better performance predictability and battery efficiency compared to existing centralized and decentralized 
platforms, while also incurring lower network traffic. 
Using both real systems and a validated simulator we show that HiveMind can scale to thousands of edge devices without sacrificing performance or efficiency, demonstrating that centralized platforms can be both scalable and performant. 


 }

\end{abstract}

%

\section{Introduction}

Swarms of edge devices are increasing in number and size~\cite{Tong16, Singhvi17, platformLab, Han19, flock18, Floreano18, mit_article, mit_article2, Almeida17, 
cncf, Markantonakis17, Faticanti18, Alfeo18, Albani17, Yuan18, Campion18, Dogar15, Nasser15, Sanneman15,Genc17,Reddi18,Faticanti18}. 
Swarms enable new applications, 
often with intermittent activity~\cite{Gan19,Lin18,Lin19,Delimitrou19,Delimitrou13,Delimitrou13d,Delimitrou13e,Delimitrou14,Delimitrou14b,Chen19,Delimitrou15,Delimitrou16,Delimitrou18,gan2018seer,Gan18b,Tong16,Singhvi17,Zhang19,Denby20,Majid20}, spanning accounting for people in disaster zones, to monitoring crops, and navigating self-driving vehicles. 
The devices themselves have low-power, modest resources, and are prone to unreliable network connections. 

As swarm sizes increase, designing a hardware-software system stack that enables programmable and performant operation for resources that span cloud and edge devices becomes a pressing need. Prior work has explored both centralized~\cite{platformLab,Alfeo18,Tong16} 
and distributed~\cite{flock18,Almeida17,clonecloud,Lin19,Denby20} approaches. In centralized systems all control, i.e., decision making on task allocation, as well as task execution happens in a backend cloud infrastructure. In distributed settings each edge device is mostly-autonomous, i.e., selects tasks to execute, and executes them locally, only transferring its output to a backend system. Neither approach is optimal. Centralized systems, while they enjoy global visibility into the swarm's state and can leverage cloud resources, quickly hit scalability bottlenecks with more edge devices. Distributed systems, on the other hand, scale better, but are hindered by the lack of coordination between devices, especially when there is redundant computation, or computation that would benefit from swarm-wide learning. 

We present \textit{HiveMind}, the first hardware-software system stack for swarm coordination that effectively bridges the gap between centralized and distributed coordination. HiveMind is a focuses on performance predictability, resource efficiency, and programmability. It relies on three key design components. First, it proposes a high-level declarative programming model for users to express the task graph of their applications, abstracting away the complexity of explicitly managing cloud and edge resources. It then uses program synthesis to explore different task placements strategies between cloud and edge, transparently to the user. Second, HiveMind uses a centralized, cloud-residing controller with global visibility into the swarm's state and available resources. To make centralized coordination scalable, HiveMind leverages serverless computing~\cite{openwhisk,lambda,azure_functions,google_functions,openlambda,openlambda2,fission}, which ties well with the intermittent activity and fine-grained parallelism of edge tasks, which do not warrant long-term resource reservations. It additionally automates task scheduling and straggler mitigation, lowering the bar for porting new applications to cloud-edge platforms. Third, at the hardware level, HiveMind proposes a reconfigurable, FPGA-based acceleration fabric for remote memory access and networking between edge and cloud resources and within cloud servers. By revisiting the entire system stack for edge swarms, HiveMind achieves the best of centralized and distributed coordination, while removing the burden of managing cloud-edge resources from the user. 

We build the entire end-to-end HiveMind system on a real swarm with 16 drones with a 12-machine backend server cluster. We also show HiveMind's generality in terms of edge devices by also porting it on a swarm of 14 terrestrial robots. We implement a benchmark suite comprised of a wide spectrum of edge applications, such as SLAM, image recognition, and weather analytics, as well as end-to-end 
multi-phase scenarios, locating stationary items in an area, and identifying the number of unique people in a field. We quantify the implications of serverless for IoT applications, as well as the trade-offs between centralized and distributed execution. 
Fig.~\ref{fig:intro} shows the performance and energy efficiency of HiveMind for a representative end-to-end scenario, compared to fully centralized and fully distributed systems. For the centralized system we show a setting that uses serverless (FaaS) and one that uses statically provisioned cloud resources of equal cost (IaaS). The scenario involves the drones trying to locate a set of tennis balls randomly placed in a sports field. We show results for the real 16 drones and a simulated 1000-drone swarm, using a new, validated simulator. In all cases, 
HiveMind greatly outperforms centralized and distributed coordinaion systems, both in terms of performance and energy efficiency. The difference is more dramatic for larger swarms, where centralized systems hit scalability bottlenecks. 
These results are consistent across all examined applications. 
Finally, we show that each software-hardware component in HiveMind is essential in achieving these benefits, although its mechanisms can be used independently as well, e.g., in the event where FPGA acceleration is not available in a cloud. 






\section{Swarm Coordination Design Trade-offs}
\label{sec:comparison}

We first quantify the trade-offs between centralized and distributed approaches in swarm coordination, to identify overheads that can be accelerated in hardware, and programmability bottlenecks that can be addressed through better software interfaces. All experiments are done end-to-end, on a real drone swarm, with a server cluster as the cloud backend. 

\begin{figure}
        \centering
        \begin{tabular}{cc}
                        \includegraphics[scale=0.17, viewport=105 10 700 480]{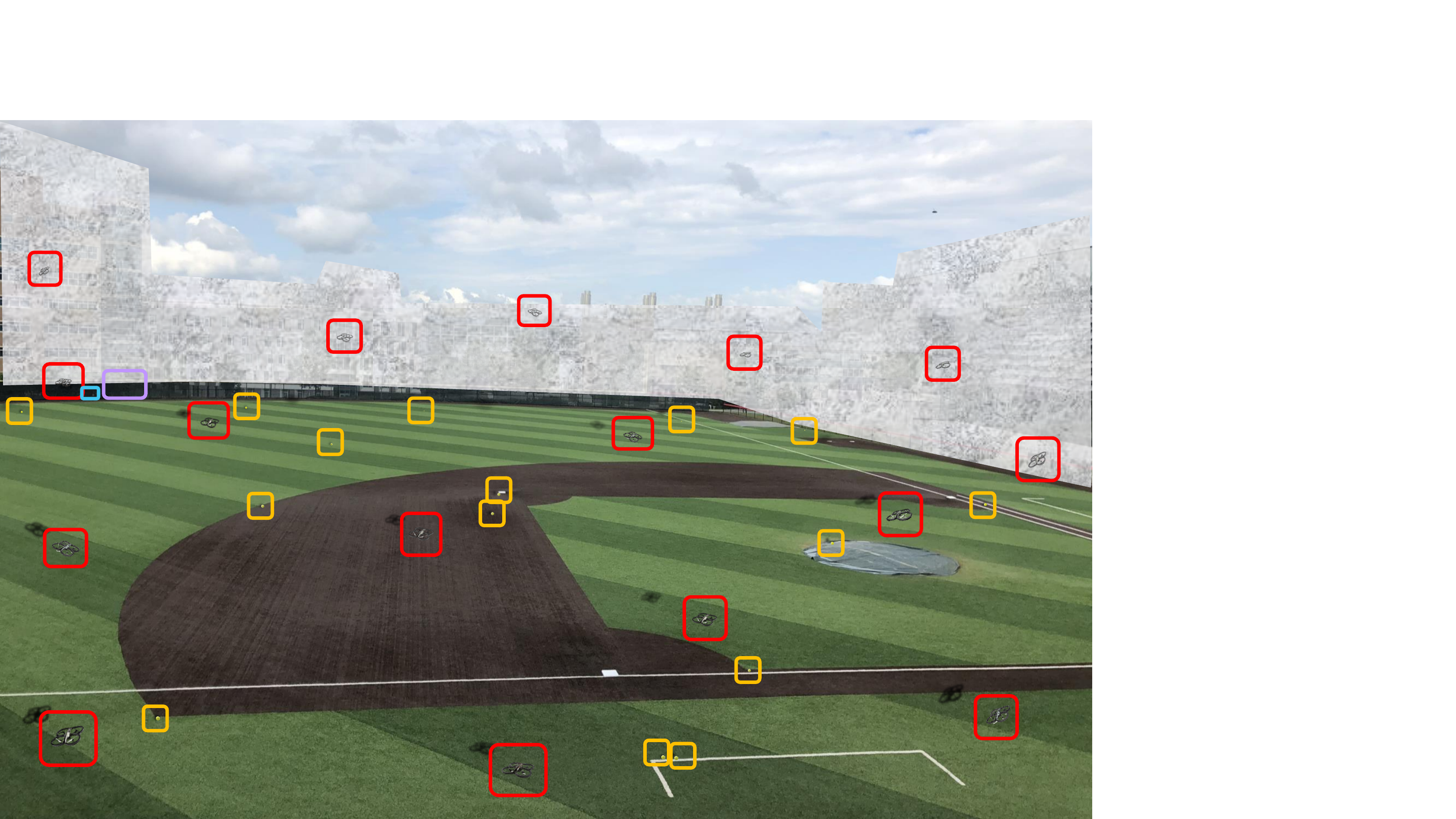} &
                \includegraphics[scale=0.145, viewport=40 10 700 480]{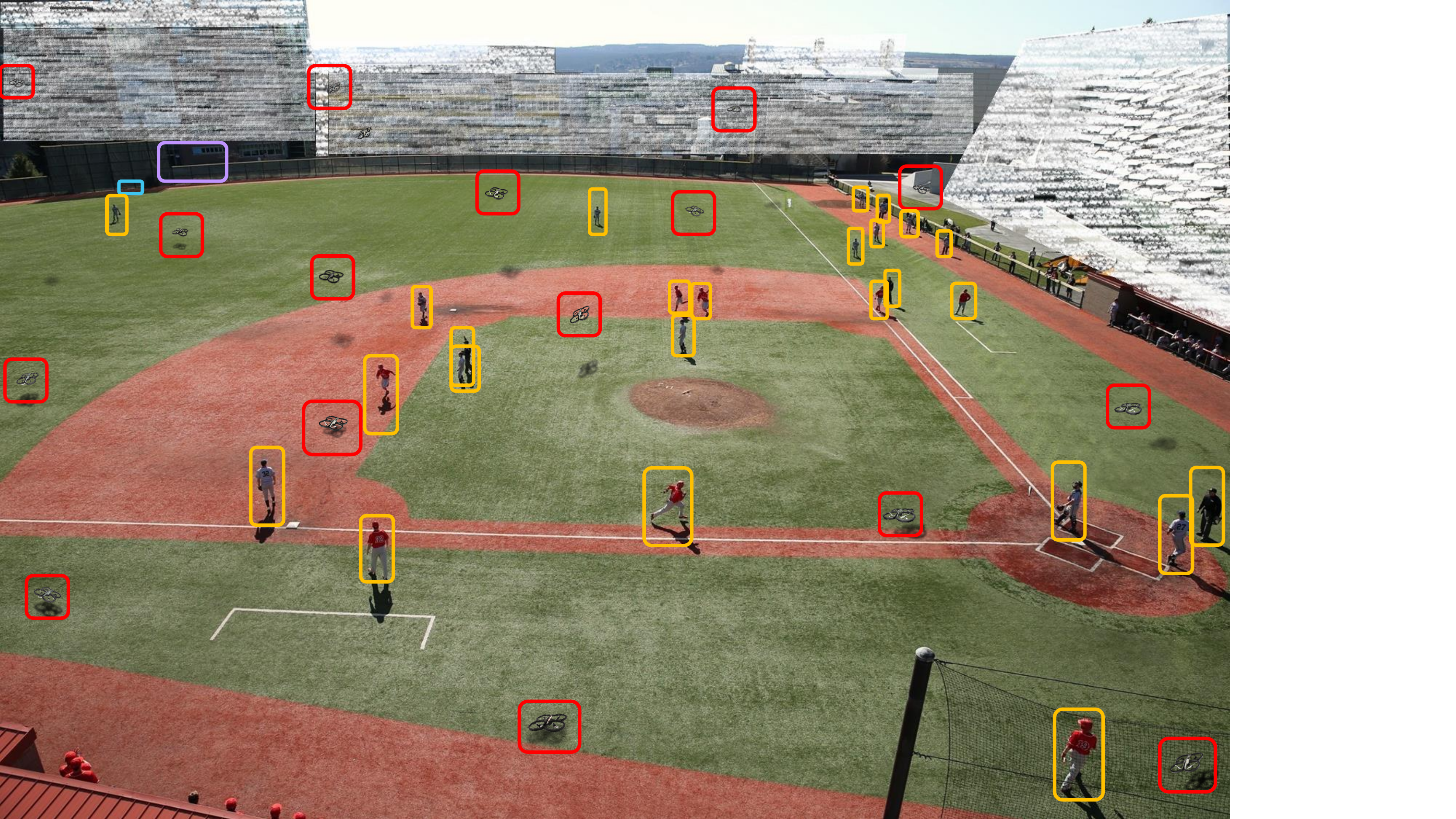} \\
        \end{tabular}
        \caption{\label{fig:scenario1} {The drone swarm executing (a) the first (static item) and (b) the second (moving people) end-to-end scenario. }}
	\vspace{-0.07in}
\end{figure}

\subsection{Methodology}
\label{sec:methodology}

\vspace{-0.04in}
\noindent{\bf{Drones: }} We use a swarm of 16 programmable Parrot {\smallcapital{AR.}} Drones 2.0~\cite{parrot}, each equipped with an {\smallcapital ARM} 32-bit Cortex {\smallcapital A8} 1GHz processor running Linux 2.6.32. There are 2GB of on-board {\smallcapital RAM}, complemented with a 32{\smallcapital GB} {\smallcapital USB} flash drive. Each drone has a vertical 720p front-camera for obstacle avoidance, and the following sensors: gyroscope, accelerometer, thermometer, magnetometer, hygrometer, and altitude ultrasound sensor. We additionally fit an 8MP camera to each drone's underside over {\smallcapital USB}, for high definition photos. Unless otherwise specified, drones collect 8 frames per second, at 2MB per frame. Drones fly at a height of $4$-$6m$ and move at $4m/s$. Their camera has a 92\degree field of view (FoV), with an approximate coverage of $6.7m\times8.75m$ per frame.

\noindent{\bf{Cluster: }}We use a dedicated cluster with 12, 2-socket, 40-core Intel servers with 128-256{\smallcapital GB} of {\smallcapital RAM} each, running Ubuntu 18.04. Each server is connected to a 40Gbps ToR switch over 10Gbe NICs. Servers communicate with the swarm with two 867Mbps LinkSys {\smallcapital AC2200} {\smallcapital MU-MIMO} wireless routers~\cite{linksys_router}. 

\noindent{\bf{Single-phase applications: }}We design and implement a benchmark suite of diverse applications, which process sensor data collected on the drones. We select both resource intensive applications better suited for cloud resources, and more lightweight services that edge devices can accommodate. These include $S1$: face recognition (identify human faces using FaceNet~\cite{facenet}), $S2$: tree recognition (identify trees using a CNN from TensorFlow's Model Zoo~\cite{tensorflow,tensorflowzoo}), 
$S3$: drone detection (detect other drones using an SVM classifier trained for the orange tag all our drones have~\cite{autonomy}), $S4$: obstacle avoidance (detect obstacles in the drone's vicinity and adjusts course to avoid them, using the obstacle detection framework in
\texttt{ardrone-autonomy}~\cite{autonomy}), $S5$: people deduplication (disambiguate between faces using FaceNet~\cite{facenet}), $S6$: maze (navigate through a walled maze using the Wall Follower algorithm~\cite{wall_follower,wall_follower2}), $S7$: weather analytics (weather prediction based on temperature and humidity levels in sensor data), $S8$: soil analytics (estimation of soil hydration from images and humidity sensor), $S9$: text recognition (image to text conversion of signs), and finally $S10$: simultaneous localization and mapping ({\smallcapital SLAM}, using image and sensor data)~\cite{slam_code}. We evaluate one service at a time to eliminate interference, however, the platform supports multi-tenancy. 

\noindent{\bf Multi-phase scenarios: }In addition to the previous applications, we also build two end-to-end scenarios, each with multiple 
phases of computation and I/O, to examine more representative use cases for drone swarms, shown in Fig.~\ref{fig:scenario1}. 
\begin{itemize}[leftmargin=*]
\item \noindent{\bf{Scenario A -- Stationary Items: }} The swarm is tasked with locating 
15 tennis balls placed in a baseball field. At time zero, the field is divided equally among the drones. Routes within each region are derived using A$^*$~\cite{astar}, where each drone tries to minimize the total distance traveled. In addition to collecting images, each drone also runs an on-board obstacle avoidance engine, based on the SVM classifier in \texttt{ardrone-autonomy}~\cite{autonomy,cylon} trained on trees, people, drones, and buildings. 
Obstacle avoidance always runs on-board to avoid catastrophic failures due to long network delays with the cloud. We have ensured that it does not have a large impact on power consumption. 

\item \noindent{\bf{Scenario B -- Moving People: }}The swarm needs to recognize and count a total of 25 people on a baseball field; the number of people is not known to the system. People are allowed to move within the field, therefore the same person may be photographed by multiple drones, requiring disambiguation. We implement two versions of the person recognition model based on the Tensorflow Detection Model Zoo~\cite{tensorflow,tensorflowzoo,coco_dataset}, one that can run on a serverless framework, and a native implementation for the edge. People disambiguation is based on the FaceNet~\cite{facenet} face recognition framework, which uses a CNN to learn a mapping between faces and a compact Euclidean space, 
where distances correspond to an indication of face similarity. 
\end{itemize}

\begin{figure}
\centering
\begin{tabular}{cc}
	\multicolumn{2}{c}{\includegraphics[scale=0.23, viewport=224 -12 1070 50]{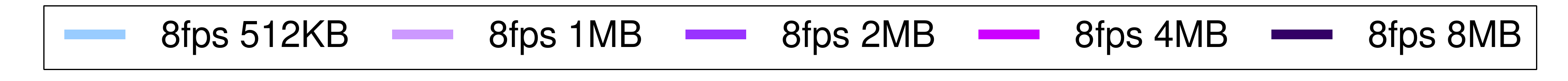}} \\
	\includegraphics[scale=0.19, viewport=44 40 840 330]{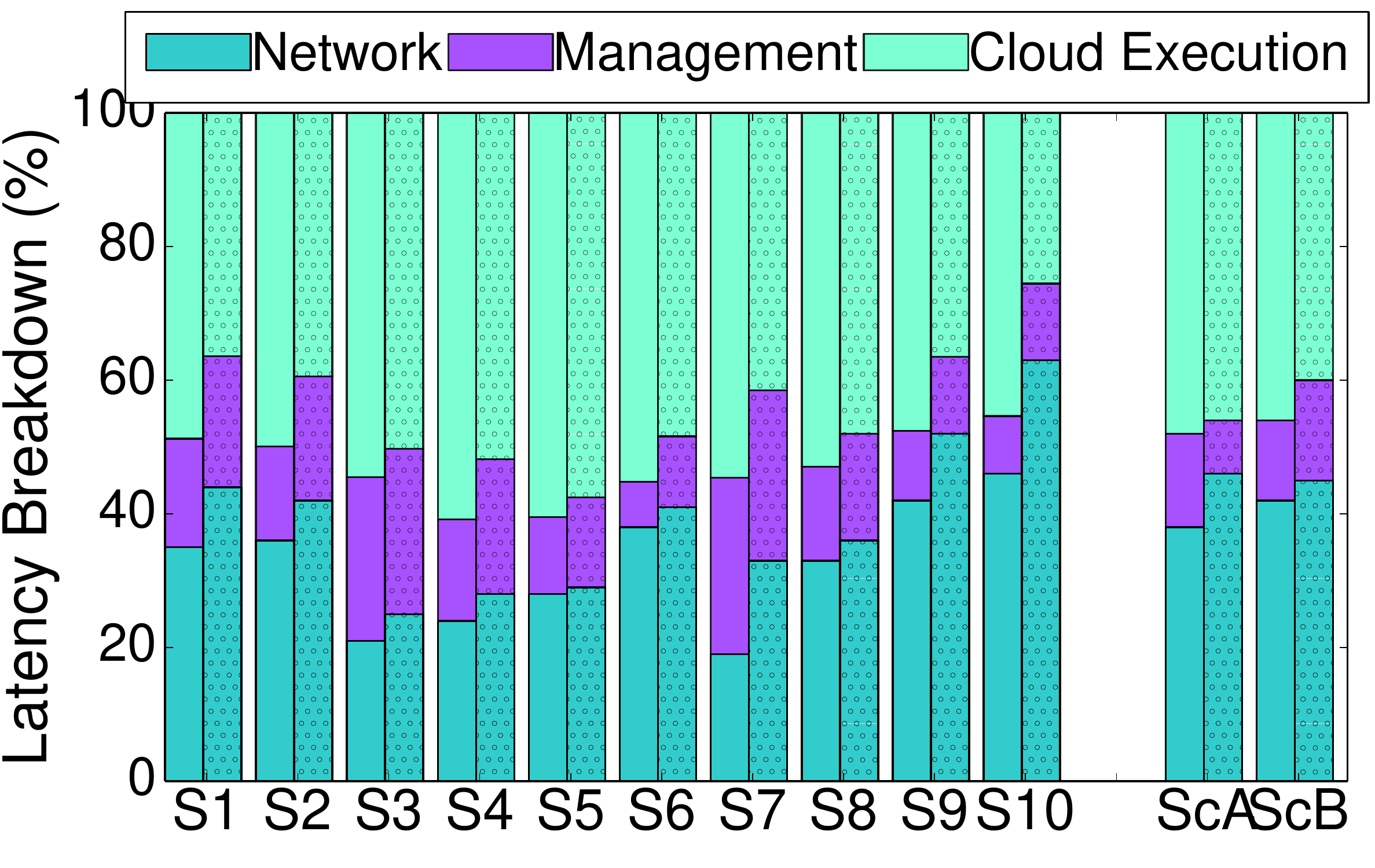} & 
	\includegraphics[scale=0.20, viewport=184 50 750 395]{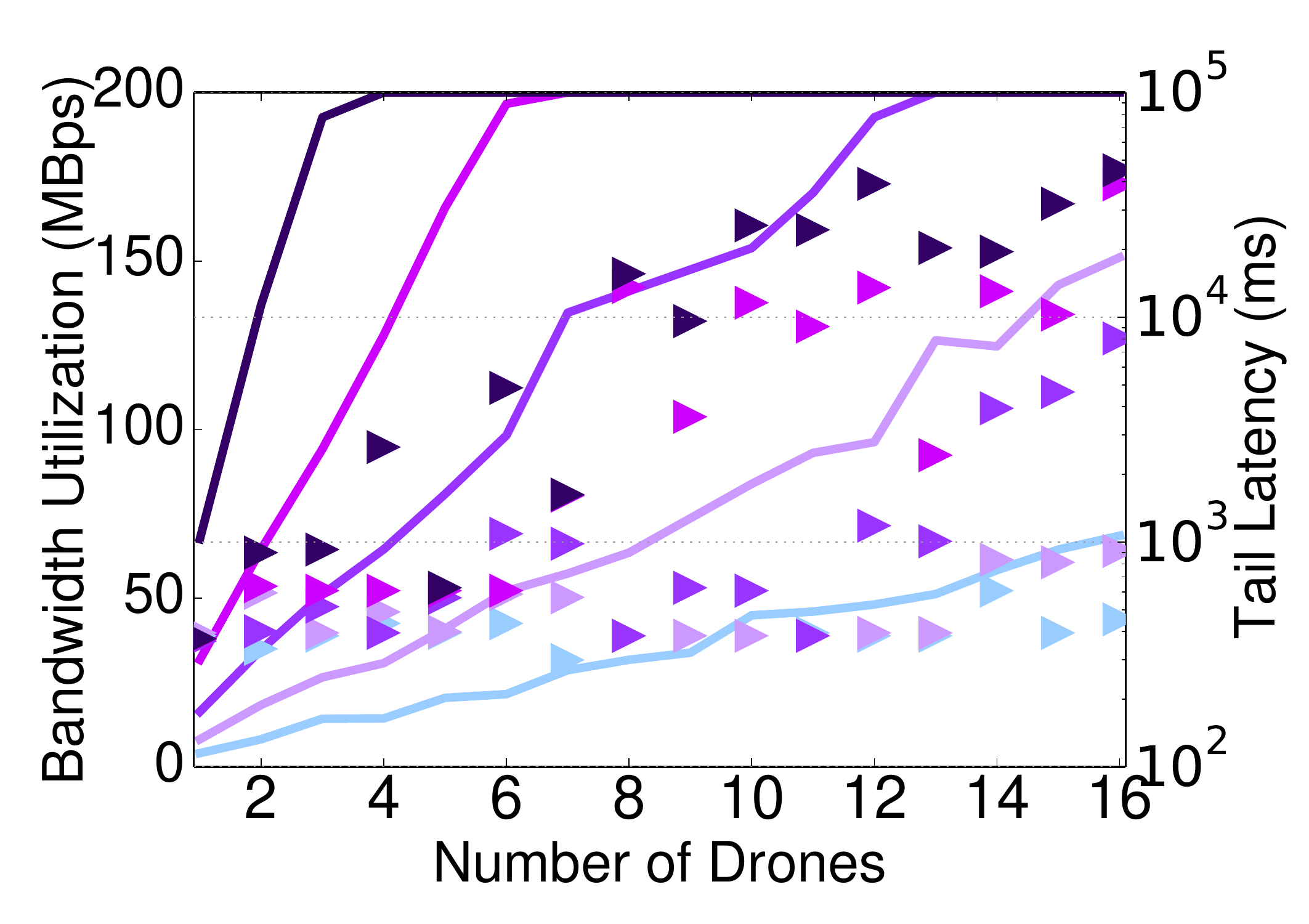} \\
\end{tabular}
	\caption{\label{fig:implications_net} (a) Network overheads (non-shaded$\rightarrow$med, shaded$\rightarrow$tail latency), and (b) bandwidth for $S1$. }
	\vspace{-0.14in}
\end{figure}

\vspace{-0.02in}
\subsection{Network Overheads}
\vspace{-0.04in}

We first examine a setting where all computation happens in the cloud, contributing to network congestion. Fig.~\ref{fig:implications_net}a shows the fraction of end-to-end latency corresponding to network processing, task execution, and management operations (task instantiation, scheduling, etc.) across the ten jobs and two end-to-end scenarios. Execution includes computation and data sharing between functions. 
Non-shaded bars show median latency, and shaded bars tail latency (99$^{th}$ pctl). Across all jobs, networking accounts for at least 22\% of median latency (33\% on average), and a higher fraction of tail latency. This is even more pronounced for the end-to-end scenarios, which involve multiple phases of computation and communication between the drones and cluster. Services are not running at max load here, so the network links are not oversubscribed. We now examine the system's scalability as network load increases. Fig.~\ref{fig:implications_net}b shows the bandwidth and tail latency for Face Recognition, as the number of drones collecting images of higher resolution increases. Tail latency remains low for fewer than 4 drones, even for max resolution (8MP). As the drones increase, the network saturates and latency increases dramatically. While the max resolution is not always required, offloading all data to the cloud limits the number of devices the framework can reliably support, or requires aggressive acceleration of networking to accommodate larger swarms. 

\begin{figure}
\centering
\begin{tabular}{ccc}
	\multicolumn{2}{c}{\includegraphics[scale=0.19, viewport=244 -50 940 50]{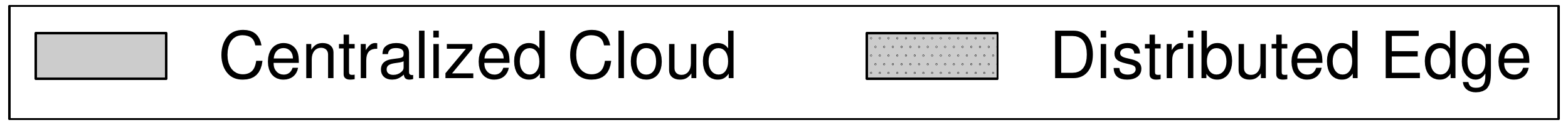}} & \\
	\includegraphics[scale=0.21, viewport=24 40 750 360]{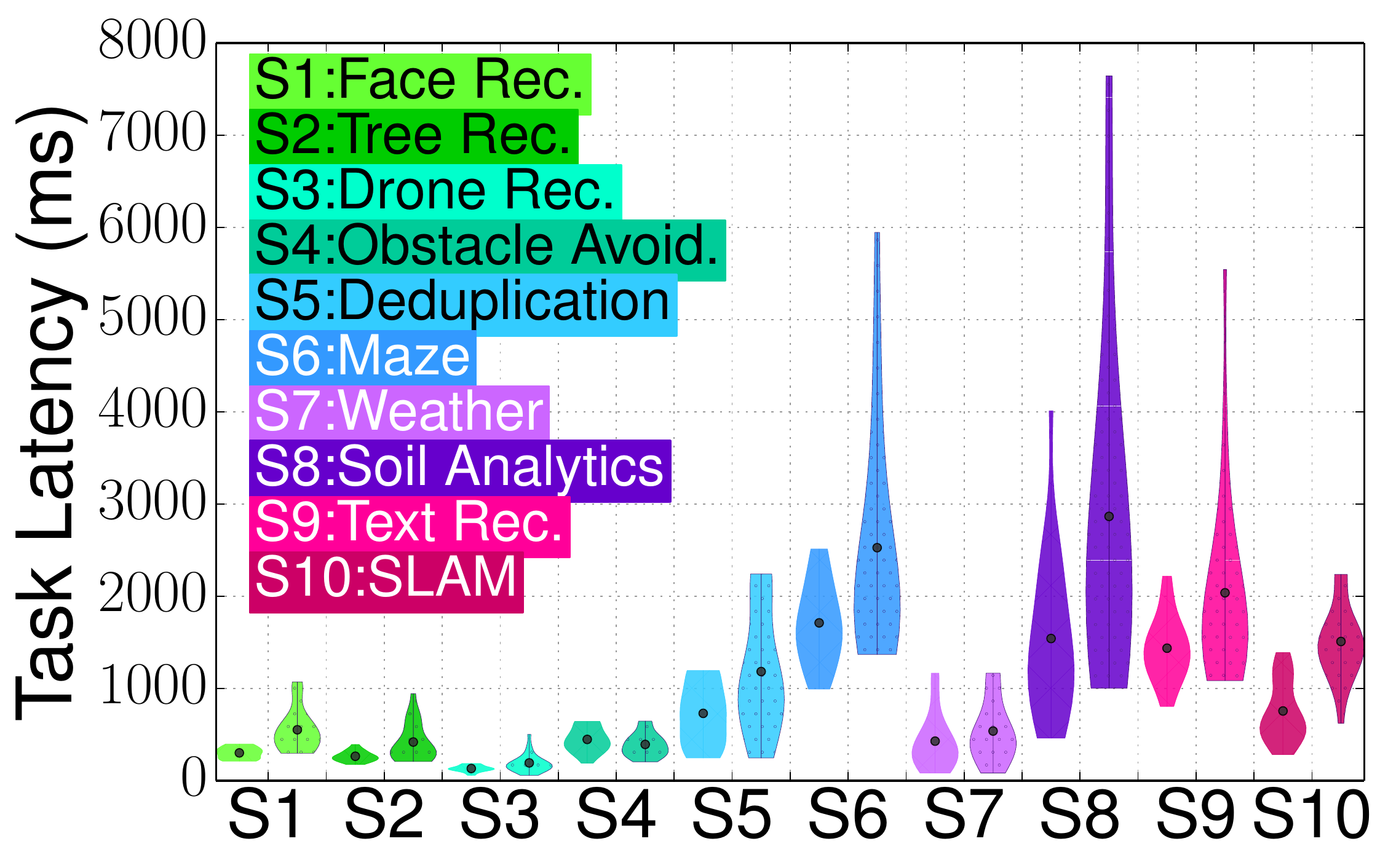} & 
	\includegraphics[scale=0.21, viewport=124 40 500 360]{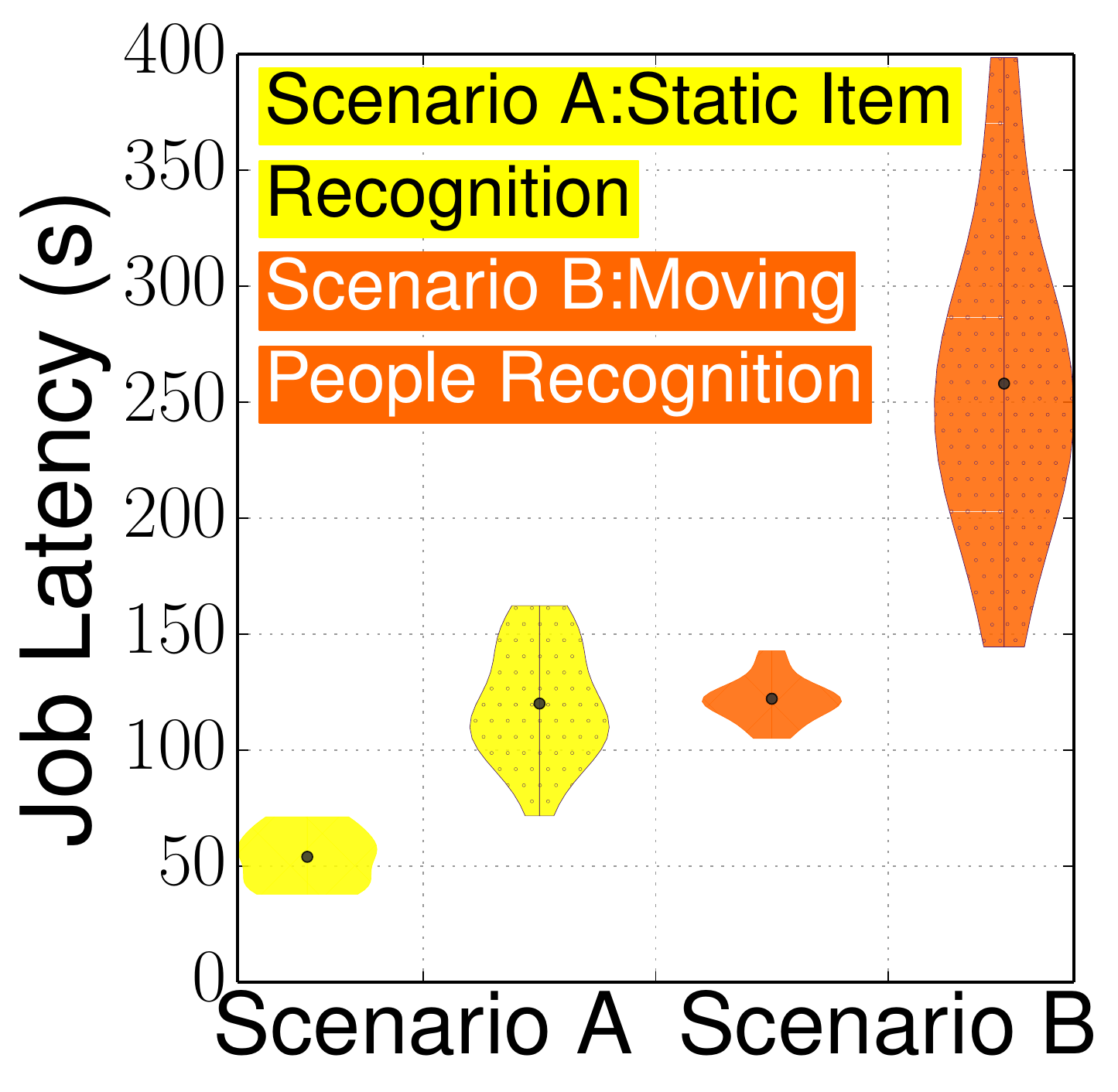} \\
\end{tabular}
	\caption{\label{fig:implications_comparison} Centralized vs. distributed execution across the (a) ten jobs and (b) two scenarios. }
	\vspace{-0.08in}
\end{figure}

\vspace{-0.02in}
\subsection{Centralized vs. Distributed Execution}
\label{sec:centr_distr}
\vspace{-0.04in}

Finally, we examine the trade-offs between fully centralized execution, where all computation and data aggregation happens in the server cluster, compared to a fully decentralized environment, where computation happens on the edge devices, and only the final outputs are transmitted to the cloud. 

To control the placement and instantiation of short-lived tasks in the backend cloud we use Apache OpenWhisk~\cite{openwhisk}, a widely-used open-source serverless framework, and the backbone of {\smallcapital IBM} Cloud Functions~\cite{shahrad19,core19,Gan19,Delimitrou19,Gan18}. OpenWhisk instantiates functions in Docker containers, via an {\smallcapital HTTP} request to an {\smallcapital{NGINX}} front-end, which triggers the OpenWhisk Controller to check a database~\cite{couchdb} for authentication, and select an Invoker to instantiate the function. Applications are in a language OpenWhisk supports (Python, node.js, Ruby, PHP, Scala, Java, Go). 
While the trade-offs between centralized and distributed execution have been explored for traditional cloud environments~\cite{shurman17,Mauro16,Lin18,Lin19}, the opportunities and challenges serverless introduces impact prior findings. We discuss these implications in more detail in Section~\ref{sec:motivation}. 
Each of the ten jobs runs for 120s, repeated 10 times, and each end-to-end scenario
runs to completion, repeated 50 times. 

Fig.~\ref{fig:implications_comparison}a shows the task latency distribution across the ten jobs, and Fig.~\ref{fig:implications_comparison}b across the two end-to-end scenarios. 
For most jobs, centralized execution achieves better and more predictable performance than 
on-board execution, despite the increased overheads from offloading data to the cloud. The difference stems both from the higher compute and memory capabilities of server nodes, and from the higher concurrency serverless can exploit compared to on-board execution. 

Three exceptions are drone detection ($S3$) and weather analytics ($S7$), which behave comparably on the cloud and edge due to their modest resource needs, and obstacle avoidance ($S4$), which achieves better performance at the edge, by avoiding data transfers, and by adjusting its route in-place, instead of waiting for the cluster to update its route if there is an obstacle. The results are similar for the two scenarios, and more pronounced for the more computationally-intensive Scenario B. Despite the performance advantage of the cloud, it also greatly increases network traffic, hurting scalability. 

Beyond the performance comparison, on-board execution quickly drains the drones' battery, 
leaving the second scenario incomplete due to several drones running out of power. 


\section{The Role of Serverless in Edge Swarms}
\label{sec:motivation}

From the previous study, it is clear that centralized control is superior in terms of performance and resource efficiency, however, it suffers from high network overheads and scalability bottlenecks. To address in particular the second of these challenges, we leverage serverless computing, a new event-driven programming framework for services with fine-grained parallelism and intermittent activity. We first quantify the opportunities 
serverless offers as the programming model of the cloud backend, and then discuss its challenges. 

Since in this study we are exclusively interested in the trade-offs between serverless and traditional IaaS or PaaS deployments for edge applications, we only consider performance metrics within the bounds of the cloud system, i.e., latency is measured from the moment sensor data arrive to the cloud until before the response is sent to the drones. This isolates any overheads that serverless may add from the impact of network congestion between edge and cloud.

\vspace{-0.02in}
\subsection{Serverless Background}
\vspace{-0.04in}

Serverless or Function-as-a-Service (FaaS) computing has recently emerged as a cost-efficient alternative for applications with high data-level parallelism, intermittent activity, fluctuating load, and mostly 
stateless operation, for which maintaining long-running reserved resources is inefficient~\cite{lambda, google_functions, azure_functions,Hellerstein18,vaneyk18,Klimovic18,Baresi17,Lynn17,core19,openwhisk,openlambda2,Lloyd18,shahrad19,xcamera,sand18,spock,sock}. 
Serverless functions are instantiated in short-lived containers or VMs, last up to a few minutes, and containers are terminated shortly after the process completes, freeing up resources. Users are charged on a per-request basis, depending on 
the amount of CPU and memory time allocated. 

Serverless gives cloud providers better visibility into application characteristics, and reduces cloud overprovisioning~\cite{GoogleTrace,Delimitrou13,Delimitrou14,Delimitrou16,Lo15,BarrosoBook,barroso_keynote}, as users are no longer responsible for explicitly reserving allocated resources. 

Functions typically take less than a second
to spawn, and most providers allow the user to spawn thousands of concurrent functions. 
Functions are instantiated in containers or VMs but are not guaranteed a specific machine type, and are scheduled by the provider to improve utilization. Serverless lowers the entry bar for new applications, but also opens them up to unpredictable performance due to function interference. Current providers only offer service level agreements (SLAs) in terms of availability, not performance, as the provider does not have visibility into application-level performance metrics. 

Internet of Things (IoT) applications are particularly well suited for serverless, given their intermittent nature, and the modest amount of on-board resources, which motivates cloud offloading~\cite{edgeai,Sarkar20,Trilles20,PInto18,tiny20,Hall19,lambda, azure_functions, google_functions,fission,openlambda,openwhisk,openlambda2}.
Serverless also introduces caveats for edge applications, especially when they have to meet quality of service (QoS) requirements,
the violation of which can cause catastrophic failures of edge equipment. While prior work has studied the architectural implications of serverless for cloud jobs~\cite{shahrad19}, 
its implications for edge services are not clear. 

\subsection{The Opportunities of Serverless}
\vspace{-0.02in}

\begin{figure}
\centering
\begin{tabular}{cc}
	\includegraphics[scale=0.21, viewport=54 20 850 320]{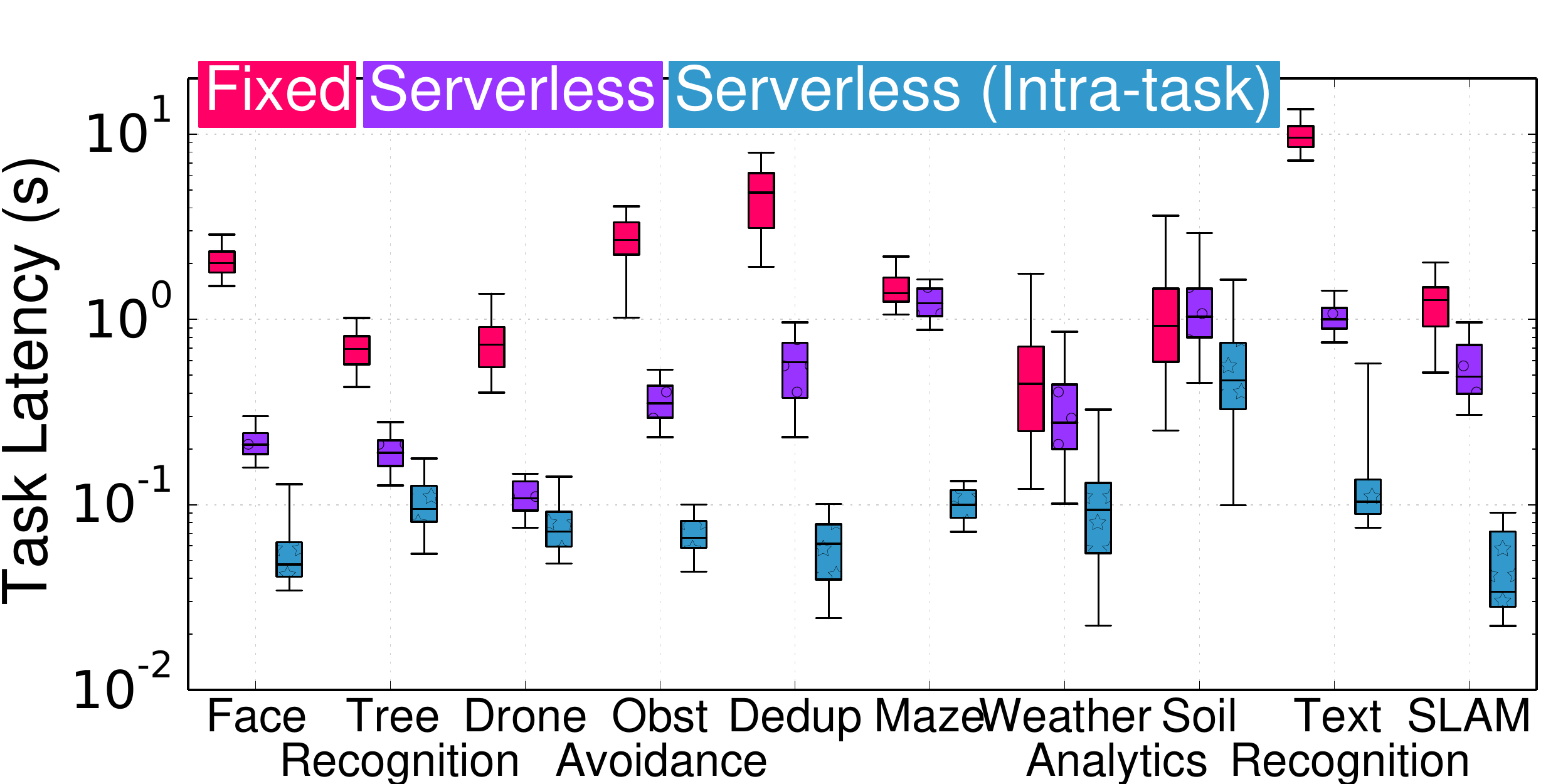} & 
	\includegraphics[scale=0.21, viewport=194 40 750 330]{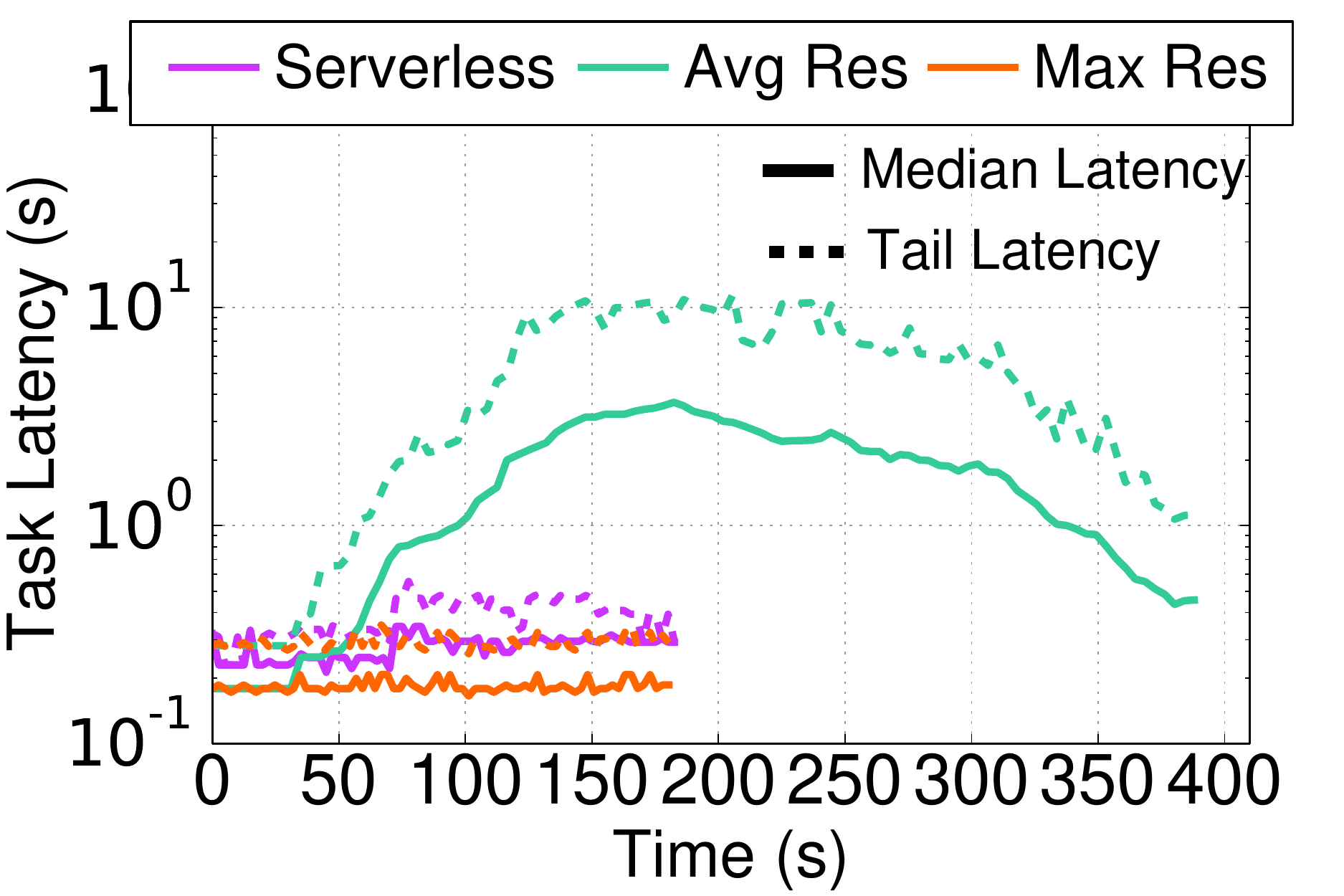} \\
\end{tabular}
	\caption{\label{fig:motivation_opportunities} {The opportunities of serverless for edge jobs: (a) Task latency with fixed and serverless deployments and 
	(b) Latency for Face Recognition with a fluctuating load. }}
	\vspace{-0.05in}
\end{figure}

\noindent{\bf{Concurrency: }}Fig.~\ref{fig:motivation_opportunities}a shows the task latency distribution across all ten applications under constant load, 
when using a fixed amount of cloud resources in a containerized (non-serverless) 
datacenter, and when using serverless without and with intra-task parallelism. An example of a task is the process of recognizing a human face in a frame batch of one second. In serverless, each task can instantiate a single function, or leverage intra-task parallelism to further improve performance. For fairness, the total amount of CPU time is the same for all deployments. The boundaries of the box plots show the $25^{th}$ and $75^{th}$ percentiles, the horizontal line shows the median latency, and the whiskers show $5^{th}$ and $95^{th}$ percentiles. 

Even without intra-task parallelism, serverless is almost always an order of magnitude faster than the fixed allocation. This is not surprising, given that serverless takes advantage of parallelism across tasks without being limited in the number of cores it can occupy (except if the user limit is reached; 1,000 functions by default on AWS Lambda~\cite{lambda}). The maze traversal, and the weather
and soil analytics do not significantly benefit from fine-grained parallelism. For the maze traversal, the tasks per second are lower than for the other jobs, as drones move slowly in the maze, hence the benefit from task concurrency is diminished, and for weather analytics, the amount of sensor data is modest and the computation is lightweight, so even the constrained fixed resources do not become oversubscribed. 

Enabling intra-task parallelism further improves performance. For jobs like image-to-text recognition and SLAM, the improvement 
is dramatic, as they have ample parallelism, and are CPU- and memory-intensive. 

This shows that as long as the edge application has ample parallelism (data-, task-, or request-level), serverless benefits performance. There are two caveats to this: first, the latency variability is typically higher in serverless, due to interference between functions sharing a physical node, and due to the instantiation and management overheads introduced by OpenWhisk. Second, parallelism does not come for free, as the user or cloud provider need to determine the available parallelism of a given job; distributing work and aggregating results also incurs overheads from 
data sharing and synchronization. 
Despite OpenWhisk introducing millisecond-level overheads for instantiating a container, traditional PaaS/IaaS clouds introduce 
several seconds of overheads to spin up new instances. 


\begin{figure}
\centering
\begin{tabular}{cc}
	\includegraphics[scale=0.20, viewport=164 -60 540 50]{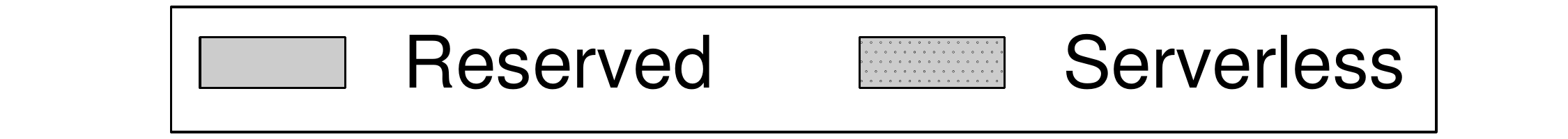} & \\
	\includegraphics[scale=0.19, viewport=44 20 750 310]{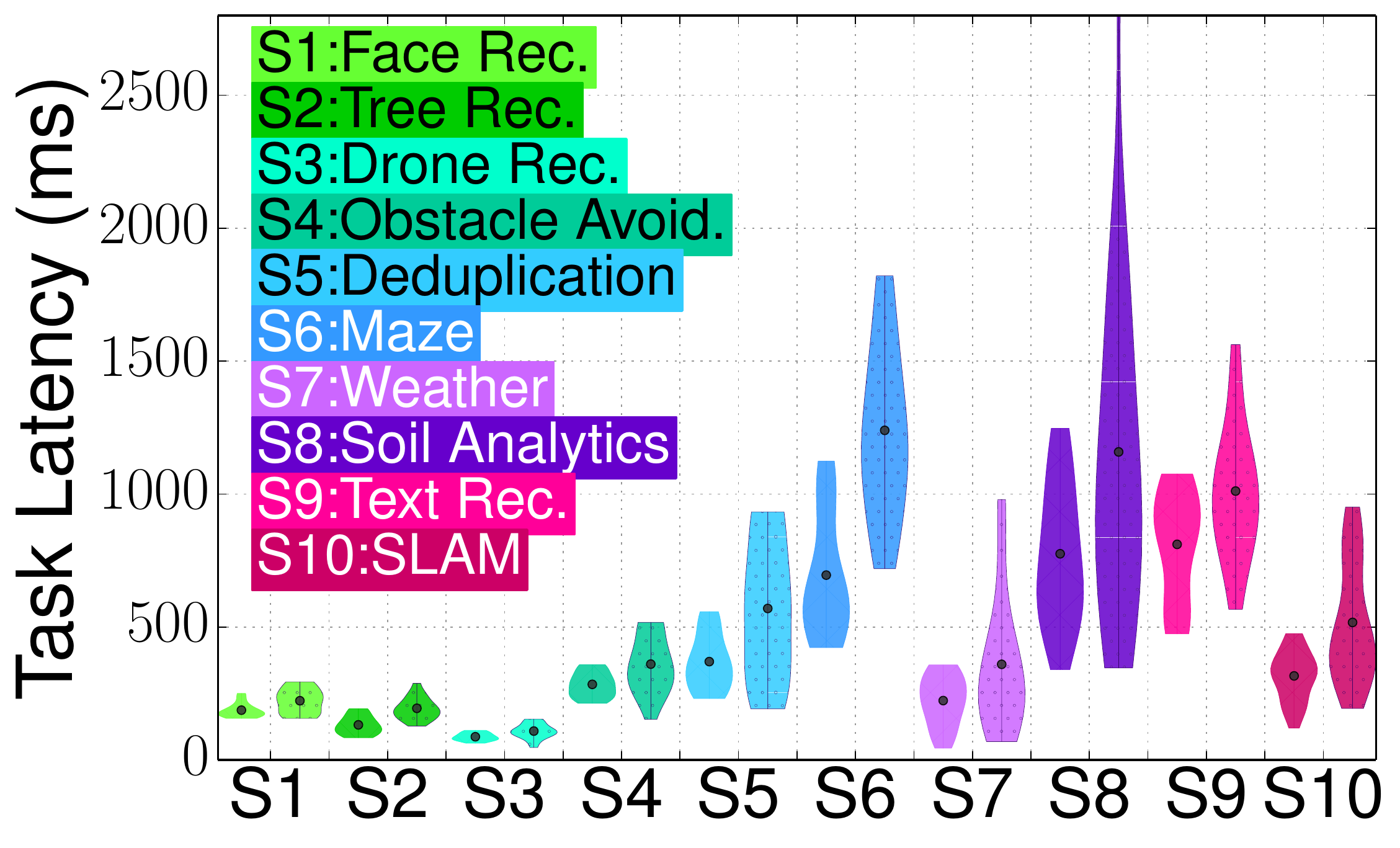} & 
	\includegraphics[scale=0.196, viewport=144 26 750 306]{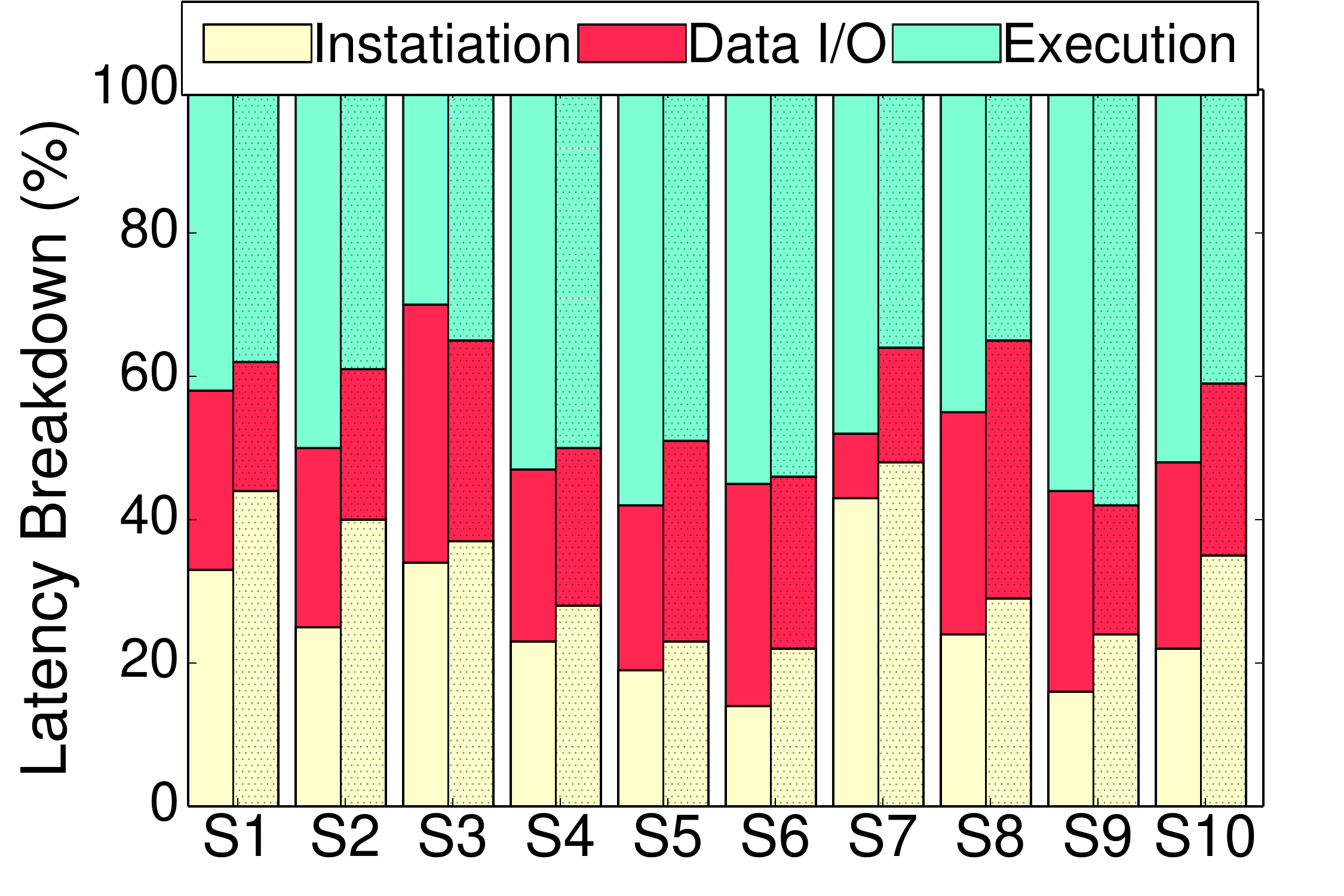} \\ 
\end{tabular}
	\caption{\label{fig:motivation_challenges} {The challenges of serverless: (a) Performance variability, (b) Impact of instantiation and data sharing. }}
	\vspace{-0.07in}
\end{figure}

\noindent{\bf{Elasticity: }}Fig.~\ref{fig:motivation_opportunities}b shows the task latency for Face Recognition ($S1$), under fluctuating load. First only one drone sends images at low rate, and progressively more drones transfer images of higher frames-per-second (fps) to the cloud. Eventually, the load decreases down to a single drone. We compare serverless to two fixed deployments; one provisioned for the average and one for the worst-case load. While serverless closely follows the load, the average-load deployment quickly becomes saturated, hurting latency. The max-load deployment also follows the load, 
but greatly underutilizes resources~\cite{GoogleTrace, Lo15, Delimitrou13, Delimitrou13d, Delimitrou15, Delimitrou14, Delimitrou16}. Given the intermittent activity of edge devices, statically provisioning cloud resources is inefficient. 





\vspace{-0.04in}
\subsection{The Challenges of Serverless}
\label{sec:motivation_challenges}
\vspace{-0.04in}

Despite its benefits, serverless introduces several challenges. 

\noindent{\bf{Performance predictability: }}
Fig.~\ref{fig:motivation_challenges}a highlights the higher performance variability serverless exhibits compared to reserved cloud resources. For all ten jobs, we show violin plots of the task latency distribution on reserved and serverless deployments. Each application runs at modest load to avoid overloading the reserved resources. Latency variability is consistently higher with serverless; a similar difference is less pronounced in the reserved resources. This is due to the instantiation overheads of serverless, which are more evident under low load when OpenWhisk terminates unused containers, the impact of OpenWhisk's scheduler when determining task placement, and the overhead of data sharing between dependent functions. OpenWhisk -- and most commercial serverless platforms -- does not permit direct communication between functions, using instead a database (CouchDB) or persistent storage for intermediate data. 

\noindent{\bf{Instantiation overheads: }}Fig.~\ref{fig:motivation_challenges}b shows the latency breakdown in serverless in terms of instantiation overheads, data sharing between functions, and useful computation. Measurements are obtained by instrumenting the OpenWhisk controller and Docker containers. Non-shaded bars show median, and shaded bars tail latency. Instantiating containers takes on average 22\% of median and 29\% of tail latency. For the short-running tasks of weather analytics that fraction exceeds 40\%, while for the more computationally-intensive maze traversal, it falls below 20\%. In all cases, it is a substantial factor to task latency, and a major contributor to tail latency. Given that edge tasks are usually short-lived, a system should minimize instantiation overheads without sacrificing resource efficiency. 

This analysis shows that while serverless has the potential to enable scalable backend computation for swarm coordination, several challenges need to be addressed, including providing a high-level programming interface for users to express their computation without needing to manage low level deployment details, like task scheduling and placement, concurrency configuration, function instantiation, and data sharing.

\section{HiveMind Design}
\label{sec:design}

Based on the analysis discussed above we design HiveMind, a scalable software-hardware system stack for the coordination of edge swarms that is programmable, performant, and resource-efficient.  
HiveMind follows a vertical design that includes a domain specific language and code synthesis toolchain, a centralized controller for a serverless cloud, and a reconfigurable acceleration fabric for remote memory access and networking. 
HiveMind abstracts away as much of the complexity of operating a cloud-edge system as possible 
from the user, without sacrificing performance or efficiency. 

To this end, HiveMind makes the following five key contributions: (1) a declarative programming model that allows users to express the high-level structure of their computation without being exposed to the complexities of deployment; 
(2) a centralized controller that automatically determines what computation should be placed in the cloud versus edge resources; 
(3) a scheduler for the serverless cloud 
that handles task placement and function instantiation; (4) a fast remote memory access accelerator that enables data exchange between dependent functions; and finally (5) a reconfigurable networking acceleration fabric that reduces the communication overheads between cloud and edge. 

\begin{figure}
       \centering
               \includegraphics[scale=0.27, viewport=340 10 600 480]{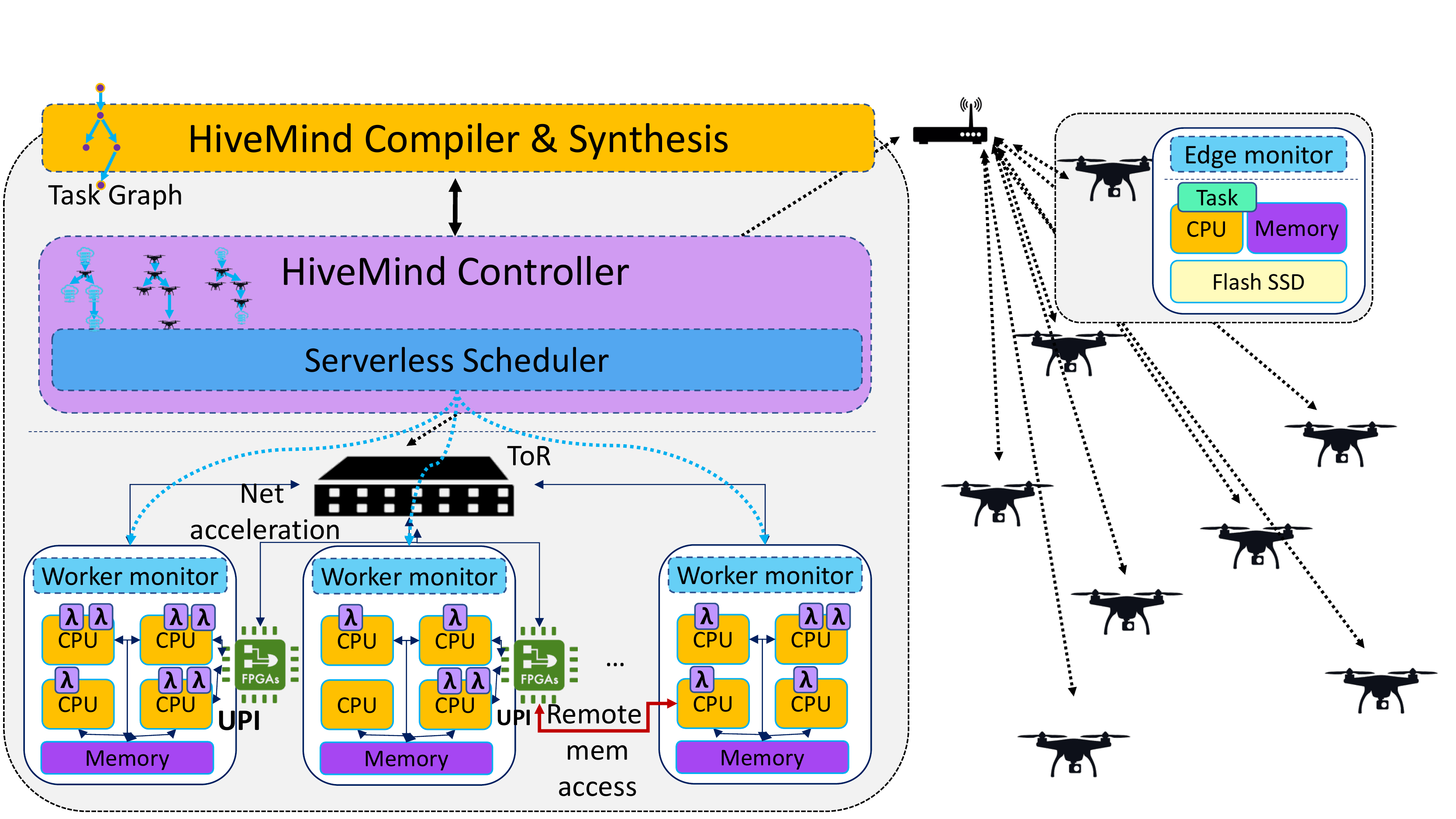}
\caption{\label{fig:hivemind_overview} {HiveMind platform overview. }}
	\vspace{-0.12in}
\end{figure}

Fig.~\ref{fig:hivemind_overview} shows an overview of the HiveMind platform. 
Below we discuss each key system component in more detail.

\subsection{HiveMind DSL and Programming Model}
\label{sec:prog_framework}

\lstdefinestyle{customcpp}{
aboveskip=0in,
belowskip=0in,
abovecaptionskip=0.08in,
belowcaptionskip=0in,
captionpos=b,
xleftmargin=\parindent,
language=Python,
morekeywords={Task, TaskGraph, Parallel, out, Stream, Restore, Schedule, Parallel, Pipeline, Serial, Synchronize, Place, Overlap, Persist, Learn, output,Isolate,args},
showstringspaces=false,
basicstyle={\linespread{0.5}\fontseries{sb}\footnotesize\ttfamily},
keywordstyle=\bfseries\color{keywordcolor},
commentstyle=\itshape\color{green!40!black},
}
Exposing all complexity associated with composing and deploying an application on a cloud-edge system to the user hinders the adoption of these platforms, or introduces performance and/or efficiency losses due to programmer faults~\cite{bugs_appengine, bugs_ops, bugs_java,bugs_dzone}. 
Instead, HiveMind follows the approach of domain specific languages (DSL)~\cite{prabhakar16,dhdl,spatial}, which abstract away most of the system complexity, 
allowing the user to focus on the high level objectives of their computation, instead of the low-level system details needed to achieve them. 

While prior work has explored task-level programming frameworks for the edge~\cite{Zhang18,azure_iot}, these systems primarily focus on querying data collected on edge devices, and/or launching single-phase, specialized computation that consumes sensor data. HiveMind is the first edge-focused DSL that targets complex, multi-phase jobs, and automates the synthesis of cross-task APIs, data sharing, and task placement. This aids programmability, as developers do not have to handle the complexity 
of defining APIs between tasks, which is a major source of bugs in such systems~\cite{Delimitrou19,Gan21,Wang17a}, especially since the supported APIs vary across edge devices. 
Public bug reports in multi-tier cloud and edge applications show that incorrect or inefficient API definition is one of the primary sources behind unpredictable performance, failures, unreachable services, or resource inefficiency~\cite{bugs_appengine,bugs_ops,bugs_dzone,bugs_java,Gan21}. 

The HiveMind DSL exposes a declarative programming interface in Python for users to express a high-level description of their task graph, and import the logic associated with different execution steps, similar to PyFlow~\cite{pyflow}, but significantly augmented to support the cross-task dependencies 
of edge applications. Listing~\ref{lst:dsl} shows a subset of the DSL's operations to define tasks and task graphs, as well as denote timing and execution dependencies between tasks. Users can specify whether tasks are allowed to run in parallel, can partially overlap, or need to execute serially. 
Listing~\ref{lst:sched} also shows some of HiveMind's optional management directives, which users can leverage to specify scheduling constraints, 
fault tolerance policies in the event of a device failure, data persistence requirements, 
as well as to enable or disable the retraining of any ML models during application execution. The two listings only show a subset of HiveMind's language; we omit data structures in the interest of space; there is support for both individual objects and data streams. 
\begin{figure}
	\begin{minipage}{0.25\textwidth}
\begin{lstlisting}[style=customcpp, caption=Subset of HiveMind's DSL operations., label=lst:dsl]
Task(name,dataIn,dataOut,code,taskArgs)
Definition of task, with 
i/o data, link to code 
path & optional arguments

TaskGraph(edgeList,constraints)
List of tasks in the 
application's control flow
and perf/cost constraints

Parallel(taski,taskj)
List of tasks a task can 
execute in parallel to 

Overlap(taski,taskj)
List of tasks a task can 
overlap with

Serial(taski,taskj)
List of tasks that cannot 
overlap

Synchronize(task,condition)
Synchronization point 
across tasks
\end{lstlisting}
	\end{minipage}
	\hspace{0.21cm}
	\begin{minipage}{0.20\textwidth}
\begin{lstlisting}[style=customcpp, caption=Optional management directives, label=lst:sched]
.
Schedule(task)
Sched. constraints & 
task priorities

Isolate(task)
The tasks requiring 
dedicated containers

Place(task)
Fix task placement 
(edge or cloud)

Restore(task)
Fault tolerance policy

Learn(task)
On/Off online learning
(one device vs. swarm)

Persist(task)
Store task's output 
in persistent storage

\end{lstlisting}
	\end{minipage}
\end{figure}
In addition to expressing the control flow of their application, users also specify the performance metrics their application must meet, in terms of execution time, latency, and/or throughput. An upper limit in terms of cost for cloud resources can also be expressed. These metrics are used by HiveMind to determine how to partition computation between cloud and edge resources. 

Listing~\ref{lst:scB} shows a simplified version of the task definitions for Scenario B, where unique people are counted in a field. 

Once a user expresses their application's task graph, the HiveMind compiler and program synthesis tool compose the end-to-end application, including automatically synthesizing the required APIs for data communication between computational steps (hereafter referred to as tiers). HiveMind generates 
two types of cross-tier APIs; one based on RPCs using Apache Thrift~\cite{thrift} for computation that may run at the edge, and another using 
OpenWhisk's function interface for tasks running on the serverless cluster~\cite{couchdb,openwhisk}. The former generates code in C++ for the cross-task APIs, while the latter uses CouchDB's communication protocol. In Sec.~\ref{sec:data_sharing} we replace OpenWhisk's default protocol with a hardware acceleration fabric for remote memory access. As the number of phases in a job 
increases, the number of APIs HiveMind needs to generate also increases. This process is entirely automated, and only happens once, at initialization. Once APIs are composed, HiveMind selects appropriate task mappings at runtime. To reduce the number of generated scenarios, 
HiveMind accepts optional hints from users regarding tasks that \textit{need} to run either at the cloud or edge, due to hardware specs, security reasons.

\vspace{0.1in}
\begin{lstlisting}[style=customcpp,frame = single, caption=Example HiveMind application for People Recognition and Deduplication. , label=lst:scB]

TaskGraph(list=['createRoute','collectImage', 
'obstacleAvoid','faceRecognition',
'deduplication'],constraint=[execTime='10s'])


Task(createRoute,inputMap,outputRoute,
'filepath/to/task/code',
load_balancer='round robin',
parentTask=None,childTask=['collectImage'])


Task(collectImage,None,sensorData,
'filepath/to/task/code',
speed='4',resolution='1024p', 
colorFormat='color',
parentTask=['createRoute'],childTask=
['obstacleAvoidance','faceRecognition'])


Task(obstacleAvoidance,sensorData,adjustRoute,
'filepath/to/task/code',
algorithm='slam',
parentTask=['collectImage']],childTask=[])


Task(faceRecognition,sensorData,recognitionStats,
'filepath/to/task/code',
trainingData='zoo',
algorithm='tensorflow_zoo',
parentTask=['collectImage']],
childTask=['deduplication'])

Task(deduplication,recognitionStats,dedupList,
'filepath/to/task/code',
sync='all',
parentTask=['faceRecognition']],
childTask=[])


Parallel(obstacleAvoidance,faceRecognition)
Serial(faceRecognition,deduplication)
Learn(faceRecognition,'Global')

Place(obstacleAvoidance,'Edge:all')
Persist(faceRecognition)
Persist(deduplication)

\end{lstlisting}

\subsection{Hybrid Execution Model}



Sec.~\ref{sec:centr_distr} showed that offloading all computation to the cloud causes network congestion, limiting scalability. On the other hand, running all tasks at the edge quickly depletes the device's battery, and is prone to unpredictable performance. 

Partitioning work between cloud and edge resources is a challenging, long-standing problem~\cite{Lin18,shurman17,Hall19,Lin19,Tong16,clonecloud}. There has been extensive work on offloading computation to a backend cloud, either by manually tagging tasks to run in the cloud, 
or by automating the offloading process, especially for mobile devices~\cite{clonecloud,Tong16,maui,odessa}.  
In the context of an edge swarm, partitioning work manually is problematic for several reasons; first, users do not have a good assessment of the performance and power consumption of different partitioning strategies, unless they can profile the application in detail. Even then, edge applications are prone to load fluctuations and failures, which can affect previous findings. 
Second, changing where computation runs affects the software infrastructure needed; for example, 
in our drone swarm, edge devices communicate with the cloud using RPCs, over TCP/IP, while cloud serverless functions communicate over OpenWhisk's CouchDB interface. Third, the use of serverless in the backend cloud changes previous trade-offs, as its instantiation overheads are higher compared to traditional cloud resources, and conversely, the ability of serverless to leverage fine-grained parallelism is more pronounced compared to long-term resource reservations. 
Additionally, prior work on automating cloud offloads~\cite{clonecloud,maui,odessa} primarily focuses on mobile devices, e.g., phones, which 
have more powerful resources than typical UAVs, do not have to also manage flight autonomy, and 
mostly optimize for power efficiency instead of performance predictability for multi-tier jobs. Similarly, prior work on offloading 
work to serverless is mostly limited to using serverless for overflow load~\cite{spock}, and still needs to address the challenges of Sec.~\ref{sec:motivation_challenges}. 

\begin{figure}
        \centering
                \includegraphics[scale=0.26, viewport=80 20 880 550]{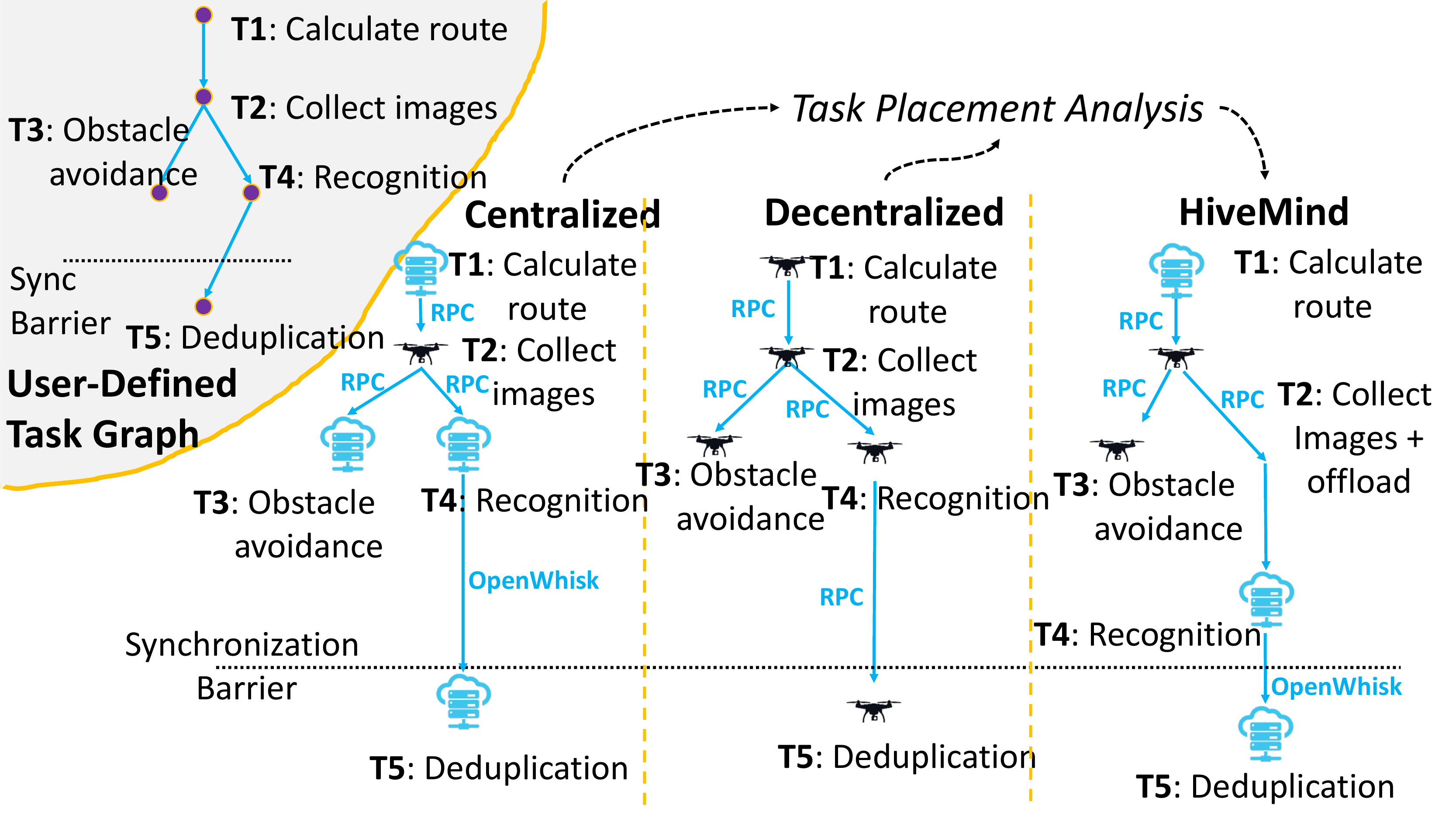}
\caption{\label{fig:hivemind_programming} {Hivemind's exploration process for task placement. }}
	\vspace{-0.08in}
\end{figure}

Fig.~\ref{fig:hivemind_programming} shows an example of this exploration for the second end-to-end scenario of Sec.~\ref{sec:methodology}. HiveMind's program synthesis then creates all  -- meaningful -- execution models, where part or all of the computation is placed on the edge devices, following any constraints provided by the user about where specific tasks should run. For a simple, 2-tier task graph ($A\rightarrow B$), HiveMind would compose the APIs for a total of 4 end-to-end scenarios 
($A_{cloud}\rightarrow B_{cloud}$, $A_{edge}\rightarrow B_{cloud}$, $A_{cloud}\rightarrow B_{edge}$, $A_{edge}\rightarrow B_{edge}$). 

Requiring the scenario to be meaningful 
reduces the search space by discarding execution models that would not make sense practically, e.g., collecting sensor data 
in the cloud. For each remaining execution model, HiveMind creates the required communication APIs, either using an RPC framework for cloud-edge communication, or using OpenWhisk's API for intra-cloud communication, and profiles the application on the target swarm. The performance and power results are presented to the user, who selects the initial work partitioning scheme that satisfies their performance 
and/or efficiency constraints. 
A user can specify constraints in terms of performance, power, cost, or a combination of these metrics. 
At runtime, HiveMind can change its task mapping if the user-provided goals are not met. Changes to task placement currently only happen at task granularity, i.e., HiveMind would not migrate a single, partially-completed task between cloud and edge at runtime. 


To maintain global visibility into all resources, HiveMind uses a centralized Controller, residing in the cluster. The controller consists of a load balancer, which partitions the available work across all devices, an interface to the scheduler 
responsible for serverless function placement, an interface to communicate to the edge devices, and a monitoring system that collects tracing information from the cloud and edge resources. 

\subsection{Serverless Cloud Scheduler}
\label{sec:scheduler}


We implement our scheduler directly in OpenWhisk's centralized controller, which is responsible for finding an appropriate Invoker to launch a function, and passing it the function information via Kafka's publish-subscribe messaging model. 
The Invoker then instantiates the function in a container~\cite{openwhisk}. While prior work has improved on native serverless schedulers to either improve resource efficiency or performance and fairness~\cite{spock,sock,salsify,xcamera,nightcore}, this work primarily targets cloud-only applications, and applications with a single or a few computation and I/O stages. 

Users can specify scheduling constraints through HiveMind's DSL, as discussed in Sec.~\ref{sec:prog_framework}. 
If scheduling constraints are specified, HiveMind's scheduler follows them, assuming they do not conflict with each other, and there are available resources in the cluster. If no scheduling constraints are specified, HiveMind places functions to optimize performance. For each server in the cluster, HiveMind deploys a \textit{worker monitor}; a lightweight process that periodically monitors the performance of active functions, and the server's utilization. Using these monitors, the scheduler identifies nodes with sufficient resources to host new functions. 

The scheduler performs two optimizations: first, in multi-tier jobs where consecutive tiers are hosted on serverless, the scheduler tries to place child functions in the same container as their parent, to avoid the costly data exchange through CouchDB. Prior work has explored co-scheduling dependent functions on the same physical node to leverage fast memory-based communication~\cite{nightcore}. This is not always possible, either because a server is overloaded, or because the child requires different software dependencies than the parent. The child may also be spawned with some delay, at which point the parent's container has been terminated. For such cases, HiveMind implements a new remote memory protocol that allows fast in-memory data exchange between functions (Sec.~\ref{sec:data_sharing}). 
In general, in serverless, once a parent task invokes a child, it terminates its own function, and OpenWhisk destroys the parent's container. In cases where descendant tasks can be placed on the container their parent used, HiveMind also places the parent's output data in a virtual memory region visible by the child, similarly to prior work~\cite{catalyzer,Babelfish,nightcore}. 

The scheduler's second optimization is to reduce instantiation overheads. In line with similar 
optimizations 
on public clouds~\cite{lambda,azure_functions,google_functions}, HiveMind does not immediately 
terminate an idling container, for the event where a new function arrives in the near future that can use it. If a new function does arrive, 
HiveMind schedules it in that container. If not, it terminates it. The amount of time an idling container remains alive is empirically set; it 
ranges between 10 and 30 seconds, given that serverless containers are instantiated much faster than traditional IaaS and PaaS containers. 

Finally, to avoid interference between functions, two containers can share a physical server, but never share a logical core. HiveMind pins containers to cores to avoid interference from the OS scheduler's decisions. For our examined applications memory contention was never an issue; however, cache partitioning and memory bandwidth partitioning can also be integrated in HiveMind for performance and security isolation. 

In Sec.~\ref{sec:scalability} we study the centralized scheduler's scalability. While the system scales to many edge devices, as with any centralized system, it can become a bottleneck. In that case, HiveMind uses multiple schedulers, each responsible for a subset of tasks, but \textit{with global visibility into all cloud and edge resources}. Such shared state cluster managers have demonstrated scalability without hurting decision quality in cloud environments~\cite{Omega13,Delimitrou15}. 


\subsection{Fast Remote Memory Access}
\label{sec:data_sharing}


As discussed above, placing a child function in the same container as its parent is not always possible. In those cases, OpenWhisk's default 
data exchange involves requests to the Controller and to CouchDB, where the results of previous functions are stored. This is expensive, 
especially when many functions try to access data concurrently~\cite{soNUMA,FASST,Nebula,rdma_1,Zsolt}. 

Instead, HiveMind uses a hardware-accelerated platform based on an Intel Broadwell FPGA-enabled architecture. 
The host is an Intel Xeon E5-2600 v4 CPU, integrated with an Arria 10 GX1150 FPGA over a UPI bus (memory interconnect). 
Incoming requests are processed 
directly by the FPGA, following an RDMA over Converged Ethernet (RoCE)-style protocol, eliminating the need to copy data between 
application memory and the data buffers in the OS, and are transferred to the CPU running the function's container via the memory interconnect. 
The physical placement of parent and child functions is known by the centralized controller. While RDMA protocols have been previously used for disaggregating resources like memory~\cite{Wang20,Tsai20,rfaas}, HiveMind is the first implementation of an FPGA-based remote memory acceleration fabric for fast communication between serverless functions. 


The FPGA-based implementation significantly improves performance by bypassing the host's network stack, and the tight integration between the host and FPGA avoids the overheads of the PCIe interfaces~\cite{rdma_1,stanford_mmio,Cambridge,dagger_cal20,IX,mtcp,soNUMA}. HiveMind directly leverages the processor's cache coherence protocol to handle dirty data tracking and demand paging with no software involvement, thus reducing the remote memory access overhead. While remote memory access deviates from serverless's default 
policy of disallowing direct communication between dependent functions, it does not break the serverless abstraction that a function can run anywhere in the cluster transparently to the user. A child function need not know the physical location of its parent (and vice versa). The child simply sees a virtualized object location for the parent's output, with address mapping handled by the FPGA. 



\subsection{Hardware-Based Networking Acceleration}

\begin{figure}
\centering
\includegraphics[scale=0.37, trim=3.8cm 1cm 1cm 3cm, clip=true]{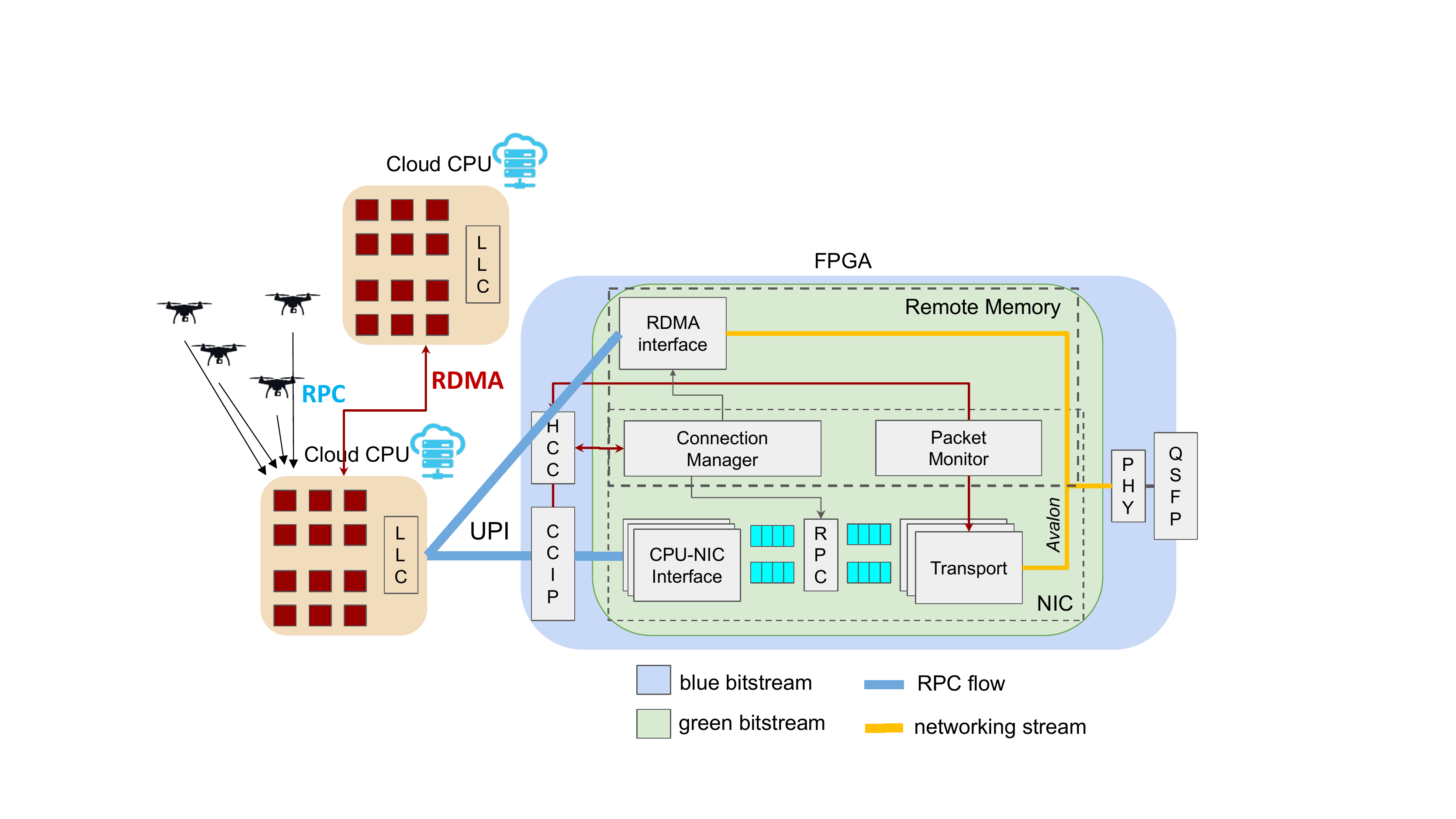}
        \caption{\label{fig:hivemind_accelerators} Overview of FPGA-based acceleration fabrics in HiveMind for remote memory access between serverless functions and RPC acceleration between cloud and edge devices. }
\end{figure}

Sec.~\ref{sec:comparison} showed that congested networks can have a dramatic impact on performance and power efficiency; similar observations have been made for cloud-based services~\cite{firestone18,stanford_mmio,catapult2,Jeyakumar13,pFABRIC,mtcp,HOMA,Zsolt,Nebula}. 
The remote memory access framework above reduces the overhead of function communication, however, accelerating traditional RPC-based networking is still required, since the edge devices use RPCs to transfer data to/from the cloud~\cite{dagger_cal20,soNUMA,RPCValet,E3,OptimusPrime,Nebula}. 
For network acceleration, we piggyback on the same hardware platform used above. Fig.~\ref{fig:hivemind_accelerators} shows 
how the tightly-coupled FPGA is connected to the host servers and the network, and partitioned between network acceleration and remote memory acceleration. We offload the entire RPC stack on the FPGA, and use the memory interconnect (UPI bus) to view the FPGA as another NUMA node, and quickly transfer data to/from the host CPU, using zero copy. Our FPGA-based implementation supports multiple threads of asynchronous (non-blocking) RPCs. The FPGA's area is large enough to support both remote memory accesses (18\% of LUTs) and RPC offloading (24\% of LUTs). We statically partition the FPGA between the two processes, although dynamic partitioning could be supported if needed. 

HiveMind's network acceleration supports a multi-connection setup between clients and servers. The networking API defines two classes: 
the \textit{RPCServer} and the \textit{RPCClient} for each client-server pair. The RPCClient encapsulates a pool of RPC 
caller threads that concurrently call remote procedures registered in the RPCServer. The software stack sets up the connections and implements a zero-copying API, 
to directly place incoming RPC requests and responses to dedicated buffers (queues) accessible by the hardware. Buffer sizes 
are configured on a per-application basis, online, through partial reconfiguration. 
The rest of the processing is handled by the FPGA NIC. When processing for a packet completes, the FPGA places the RPC payload in a shared memory buffer. 
Packets are processed to completion by a single thread. 

In contrast to existing programmable NICs, which leverage PCIe-based interfaces~\cite{firestone18,HotNet,TONIC,mtcp,IX}, we again use a NUMA memory interconnect to interface the host CPU with the FPGA, to optimize the transfer of small RPCs, which are common in edge devices. The NUMA memory interconnect is encapsulated into the CCI-P protocol stack~\cite{CCIP}. 

The intuition behind the use of an FPGA for network acceleration is that it allows the networking fabric to be reconfigurable, which suits the diverse needs of edge applications. 
Reconfiguration is split in \textit{hard} and \textit{soft} reconfiguration. The former is only used for coarse-grained control decisions, such as selecting the CPU-NIC interface protocol or the transport layer (TCP or UDP). Soft reconfiguration, on the other hand, is based on soft register files accessible by the host CPU via PCIe, and their corresponding control logic. It is used for configuring the batch size of CCI-P transfers, provisioning the transmit and receive queues, configuring the queue number and size, configuring the number of active RPC flows, and selecting a load balancing scheme. Soft reconfiguration incurs some small overheads, however, it enables fine-tuning the acceleration fabric to the needs of different applications. 

The server's NIC simply forwards packets to the FPGA without processing them in the host CPU. 
HiveMind's network acceleration achieves 2.1us round trip latencies between cloud servers connected to the same ToR switch, and a max throughput with a single CPU core of 12.4Mrps for 64B RPCs. This improves the system's performance predictability, and 
frees up a lot of CPU resources, which can be used for function execution. In Sec.~\ref{sec:evaluation} we evaluate how these benefits factor into end-to-end performance. 



\subsection{Other Features}


\noindent{\bf{Continuous learning: }}A benefit of centralized coordination is that data from all devices can be collectively used to improve the learning ability of the swarm. 
A user can enable or disable continuous learning in their application description. 
If enabled, instead of only using one device's decisions to retrain it, HiveMind aggregates the entire swarm's decisions, and retrains all devices jointly, which significantly accelerates the improvement of their decisions. 

\begin{wrapfigure}[10]{l}{0.27\textwidth}
	\centering
	        \includegraphics[scale=0.19, viewport=200 100 580 460]{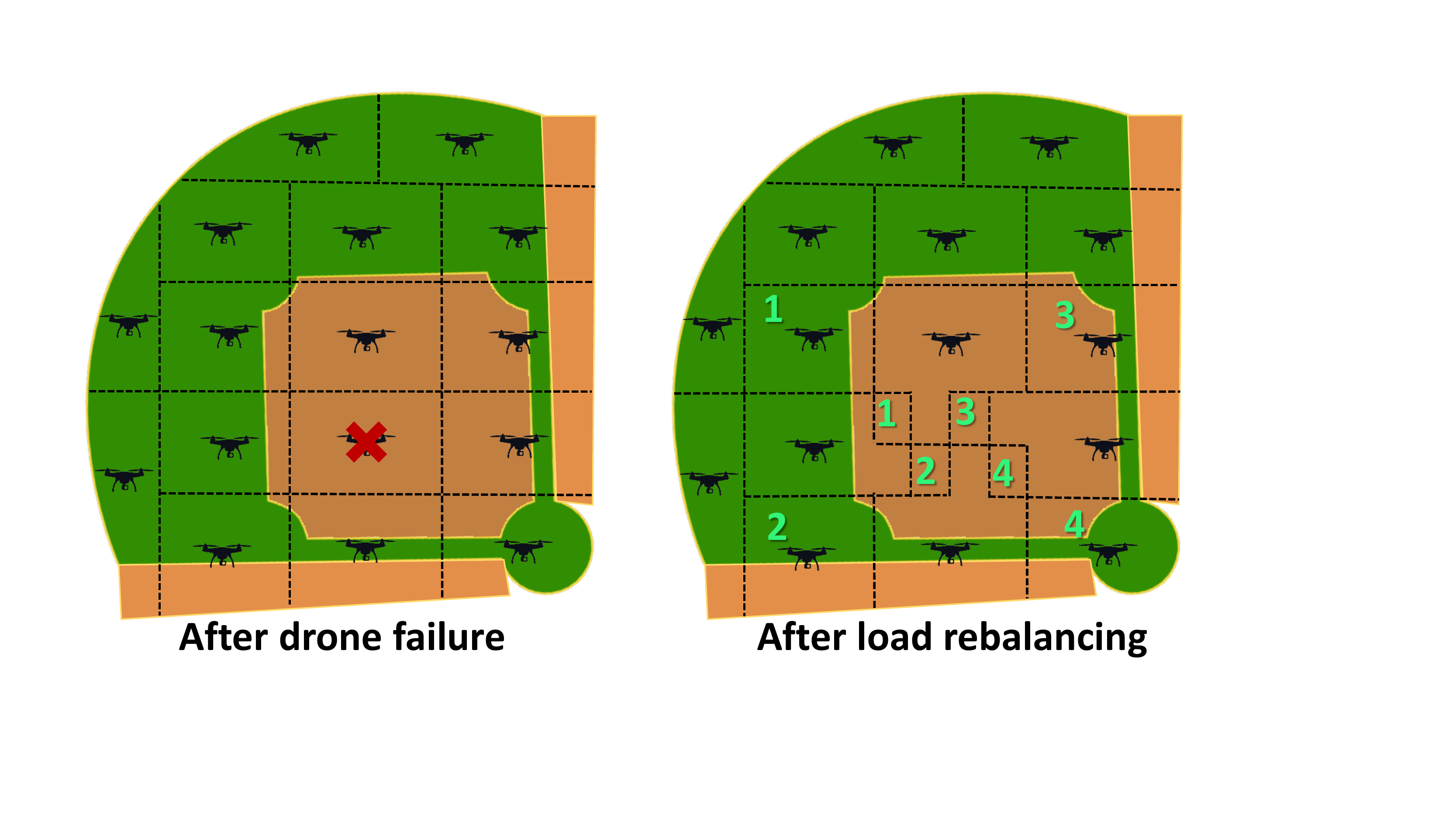}
\caption{\label{fig:fault_tolerance} {Load repartitioning to handle a drone failure. }}
\end{wrapfigure}

\noindent{\bf Fault tolerance: }Edge devices are prone to failures. All devices send a periodic heartbeat to HiveMind (once per second). If the controller does not receive a heartbeat for 
more than 3s, it assumes that the device has failed. HiveMind handles such failures by repartitioning the load among the remaining devices. Fig.~\ref{fig:fault_tolerance} shows such an example for our application scenarios. 

Immediately after HiveMind realizes that the red-marked drone has failed, it repartitions its assigned area equally among its neighboring drones assuming they have sufficient battery, and updates their routing information. Depending on which device has failed, this involves reassigning work to a variable number of devices. 
Users can also specify fault tolerance policies. 

\noindent{\bf Straggler mitigation: }HiveMind has a monitoring system that tracks function progress, and flags potential \textit{stragglers}. If a function takes longer than the 90$^{th}$ percentile of that job's functions, OpenWhisk respawns it on new servers, and uses the results of whichever task finishes first~\cite{Ousterhout13,tailatscale}. 
The exact percentile that signals a straggler can be tuned depending on the importance of a job. If several underperforming tasks all come from the 
same physical node, that server is put on probation for a few minutes until its behavior recovers. OpenWhisk also respawns any failed tasks by default. 

\begin{figure*}
        \begin{minipage}{0.35\textwidth}
        \begin{tabular}{cc}
        \multicolumn{2}{c}{\includegraphics[scale=0.204, viewport=0 5 740 30]{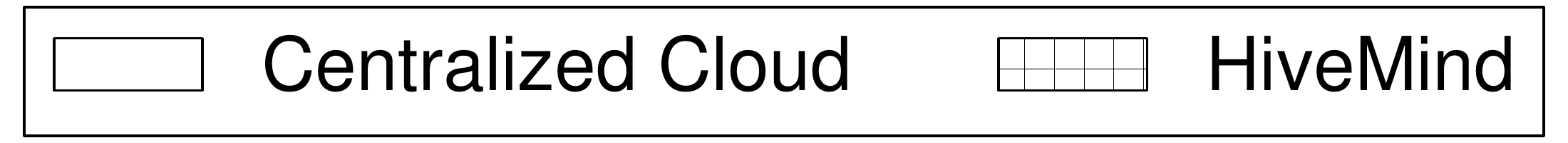}} \\
        \includegraphics[scale=0.194, viewport=60 44 860 390]{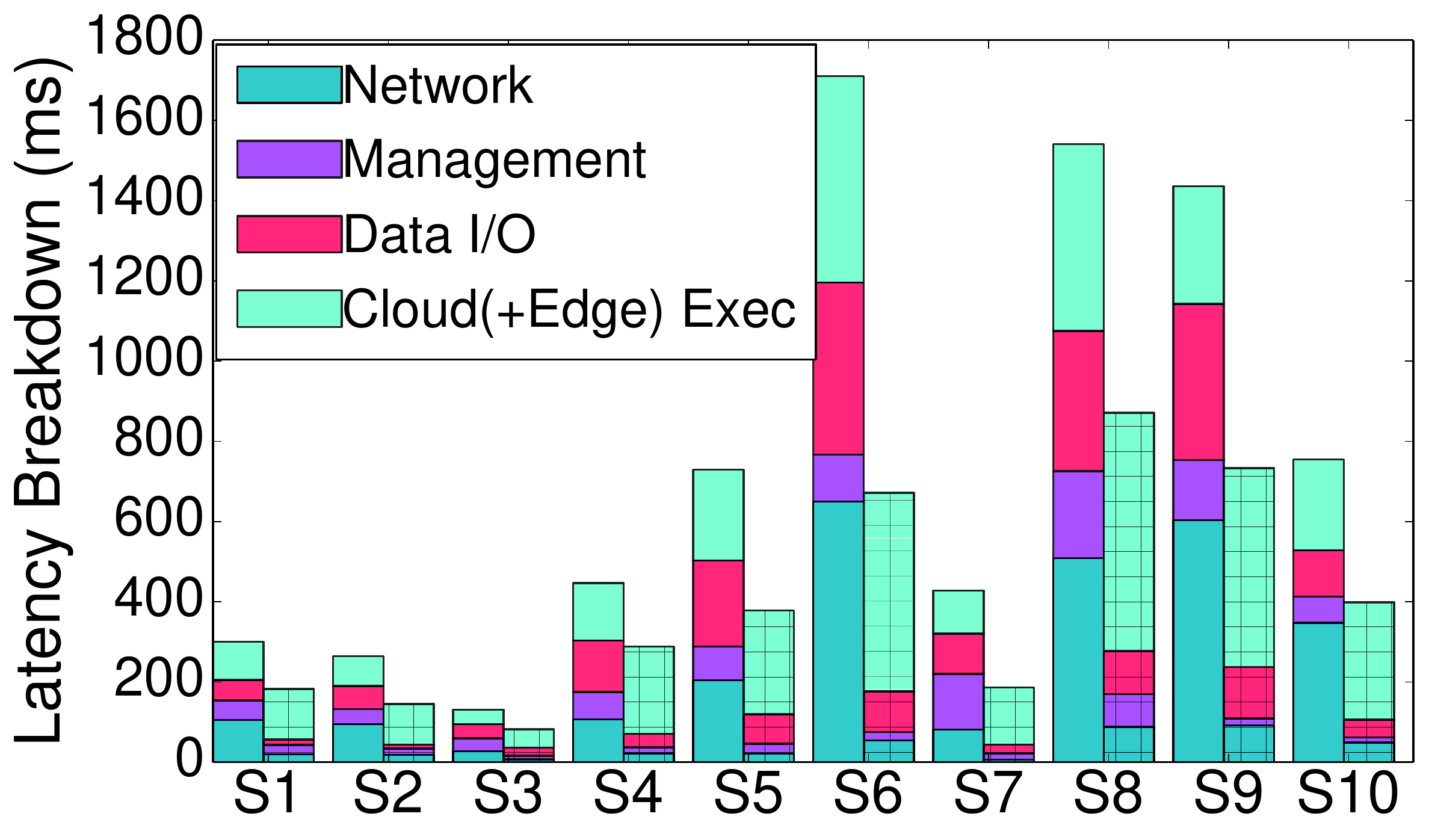} &
        \includegraphics[scale=0.194, viewport=200 44 510 390]{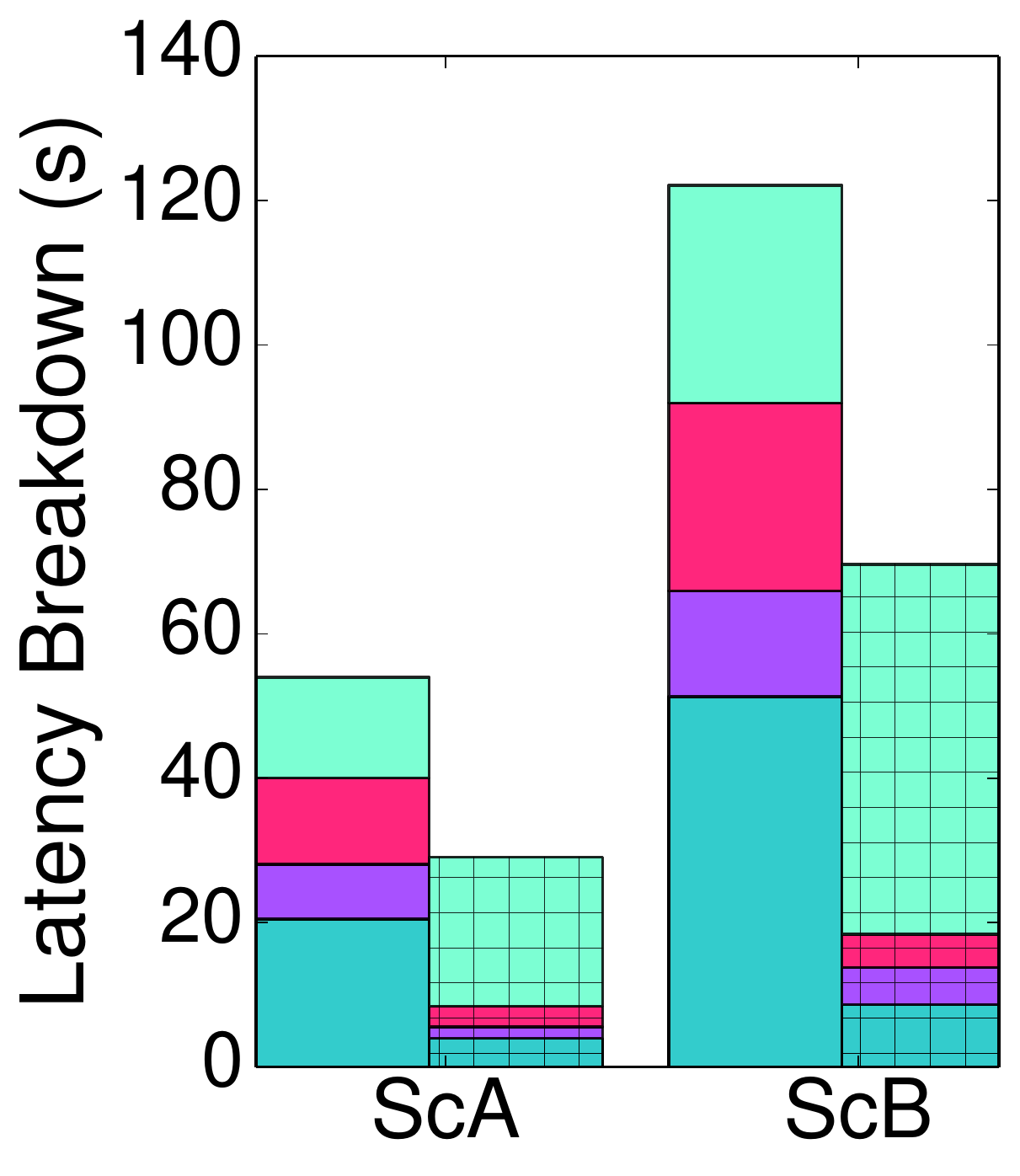} \\
        \end{tabular}
        \caption{\label{fig:breakdown_hivemind} Performance benefit breakdown. }
\vspace{-0.03in}
\end{minipage}
\hspace{0.6cm}
\begin{minipage}{0.63\textwidth}
        \begin{tabular}{cc}
                \includegraphics[scale=0.190, viewport=0 44 1320 400]{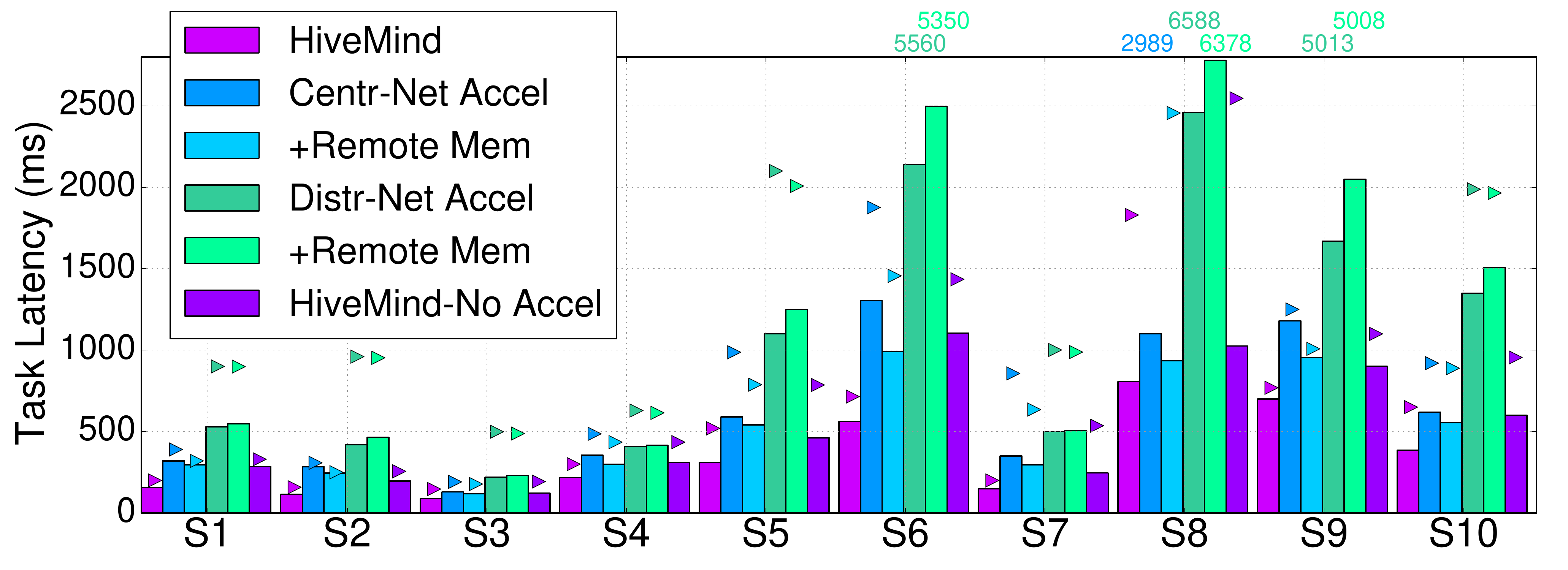} &
        \includegraphics[scale=0.190, viewport=200 44 510 400]{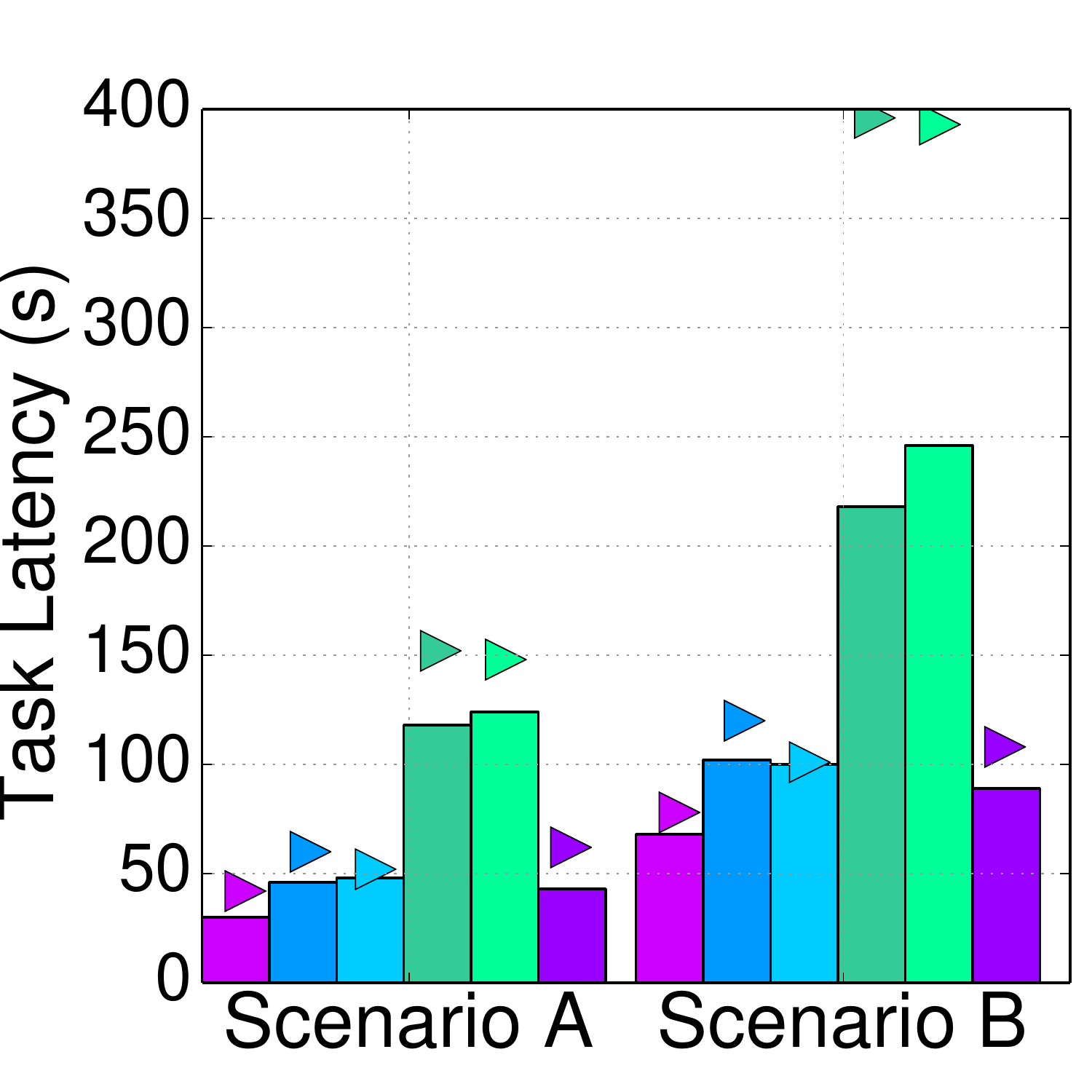} \\
        \end{tabular}
        \caption{\label{fig:incremental} Evaluation of benefits from subsets of HiveMind's techniques. }
\vspace{-0.03in}
\end{minipage}
\end{figure*}
\begin{figure}
\centering
        \begin{tabular}{cc}
        \multicolumn{2}{c}{\includegraphics[scale=0.196, viewport=220 -44 780 60]{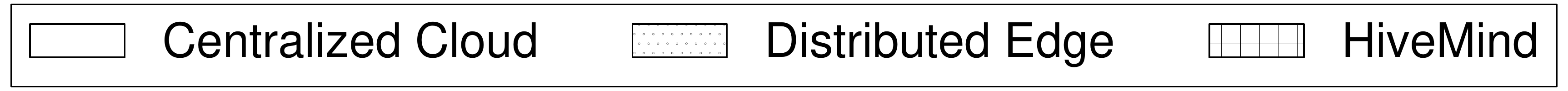}} \\
        \includegraphics[scale=0.17, viewport=110 64 1010 360]{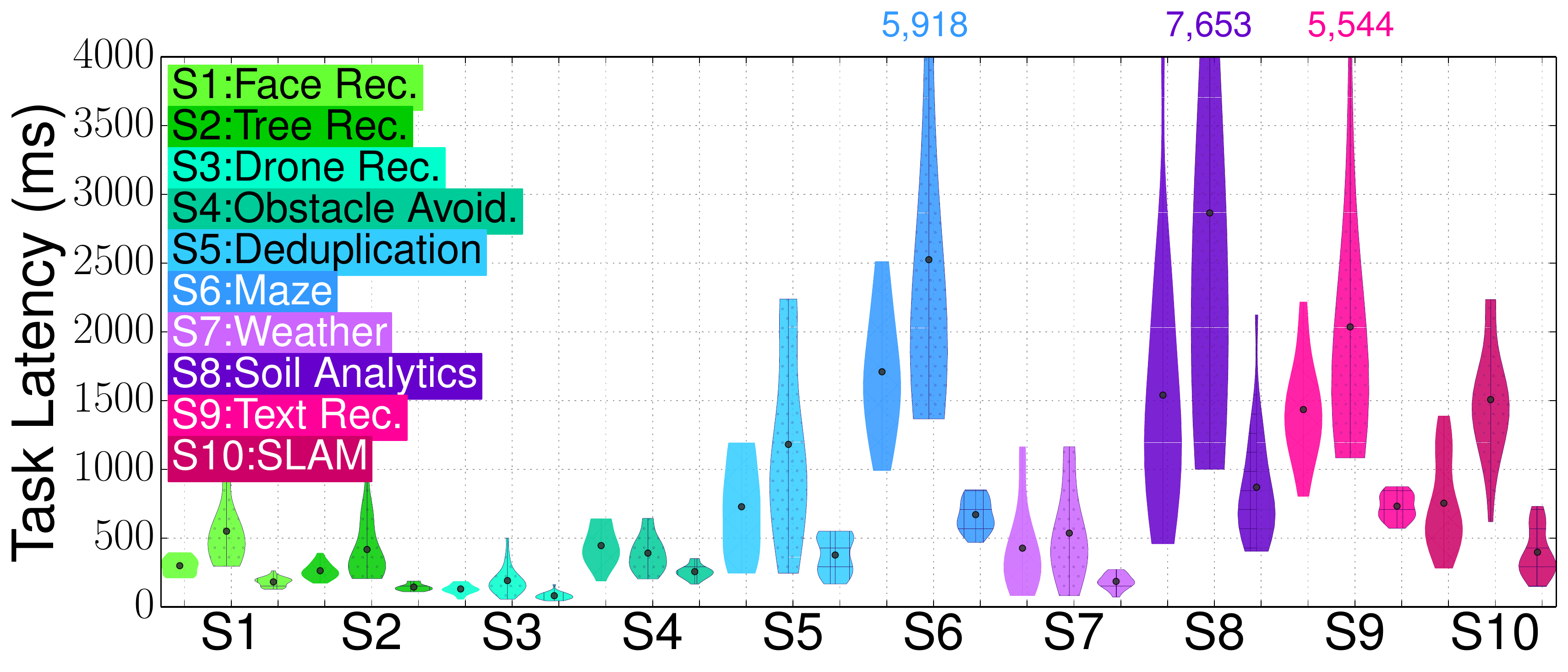} &
        \includegraphics[scale=0.17, viewport=90 64 410 360]{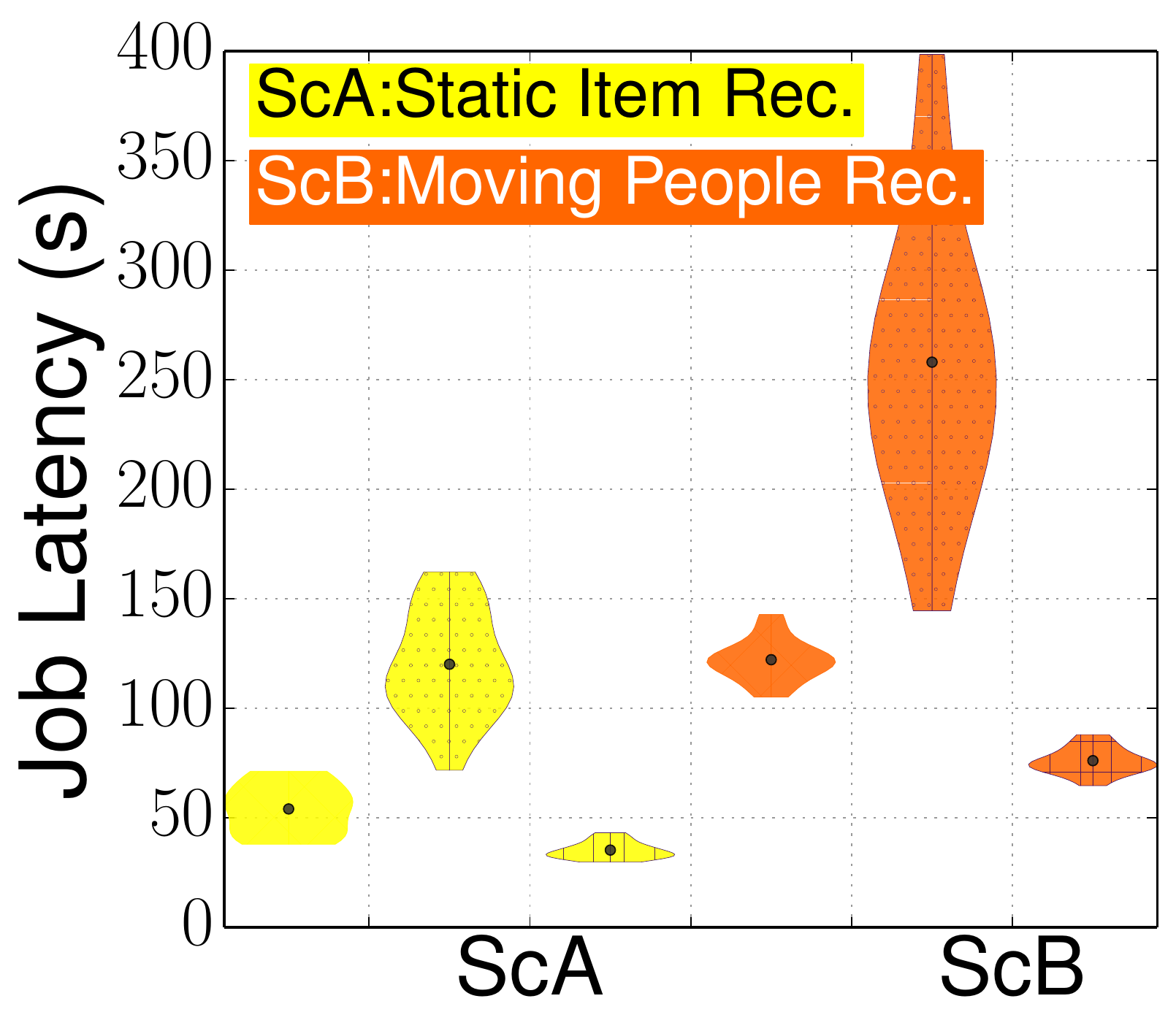} \\
        \end{tabular}
        \caption{\label{fig:latency_hivemind} Latency PDFs across applications with HiveMind. }
	\vspace{-0.03in}
\end{figure}

\subsection{Implementation}





The HiveMind compiler and program synthesis are written is \texttt{28,000} lines of C++ and Python. 
The controller is written in \texttt{18,000} lines of C++, and 
supports Ubuntu 18.04 and newer versions. 
The controller is implemented as a centralized process, 
with two hot standby copies that can take over, in case of a failure. 
We have also implemented a monitoring system that tracks application progress 
and device status, and verified that it has no meaningful impact on performance;
less than 0.1\% on tail latency, 
and less than 0.2\% on throughput. Both hardware acceleration processes (remote memory and networking) are implemented using Verilog. 
HiveMind supports applications in Python, C++, Scala, and node.js.

\vspace{-0.06in}
\subsection{Discussion}

HiveMind by default assumes that the platform has full control over cloud and edge resources to appropriately 
place functions on physical machines. However, the techniques in HiveMind are modular, and could be used separately if full system control 
is not available, as is the case in a public cloud. In such a setting, HiveMind can generate 
the low-level code of an application from its high level specification, and identify the best mapping 
between cloud and edge resources. If the cloud provider additionally has support for network-connected FPGAs, HiveMind could harness benefits 
from network and remote memory acceleration. It would, however, 
lose the advantages of controlling the physical task placement. 
Alternatively, if FPGA support is not available, HiveMind would still offer programmability and task placement benefits, but applications would be prone 
to high network overheads and overheads from the serverless framework's default data exchange protocol (CouchDB for OpenWhisk). 







\section{Evaluation}
\label{sec:evaluation}

\subsection{Performance Analysis}

We first examine HiveMind's performance predictability. 
Fig.~\ref{fig:latency_hivemind} shows the performance of all applications with 
HiveMind compared to the centralized and decentralized platforms. 
HiveMind's performance is consistently better and less variable compared 
to both other systems. The applications that benefit the most from HiveMind's design 
are compute- and memory-intensive services, like maze traversal, image-to-text recognition, and the second 
end-to-end scenario (moving people recognition) for which offloading all data to the cloud incurs high network 
overheads, and computing on the drones results in poor and unpredictable performance. 
Services like drone detection ($S3$) and obstacle avoidance ($S4$) exhibit smaller benefits, consistent 
with out findings in Sec.~\ref{sec:motivation}. 

We now examine where these benefits come from. Fig.~\ref{fig:breakdown_hivemind} shows 
the tail latency breakdown for the centralized system and HiveMind. 
Network acceleration in HiveMind has a drastic impact on latency, with time for networking dropping from 33\% on average to 9.3\%. 
Second, the time associated with management operations, such 
as container instantiation and scheduling also drops significantly. Most benefits come from HiveMind 
avoiding instantiation overheads, despite its scheduler incurring slightly higher overheads than
the default OpenWhisk Controller. Third, HiveMind's remote memory access considerably reduces data exchange latency, by avoiding 
CouchDB accesses. Finally, 
the fraction of end-to-end latency devoted to execution time in HiveMind increases compared to the centralized system. 
This is not surprising, as HiveMind maps some tasks on the slower edge devices. However, by 
eliminating all other system overheads, 
HiveMind's end-to-end 
performance is 56\% better than Centralized on average, and up to $2.85\times$, while also using less battery and bandwidth. 

Finally, we examine the incremental benefits of the techniques in HiveMind. 
Specifically, we compare HiveMind to a centralized system with network acceleration, one that also has remote memory access acceleration, 
distributed systems with one or both types of acceleration, and Hivemind with no acceleration but hybrid execution. Fig.~\ref{fig:incremental} shows the comparison in terms of median (bars) and tail latency (markers) for all ten jobs and two end-to-end scenarios. In the case of the centralized system with network acceleration, performance benefits from improving networking between edge and cloud resources, although still remains far from HiveMind. Even when remote memory access acceleration is enabled, HiveMind still outperforms the centralized system, as it does not overload cloud resources, by allowing some edge computation. On the other hand, the distributed system barely benefits from hardware acceleration, as most computation happens at the edge, and only final results are transferred to the cloud. Finally, HiveMind without acceleration still benefits from hybrid execution, but is prone to high networking and data transfer overheads. Overall, this analysis shows that no single technique in HiveMind is sufficient to address the performance and power requirements of edge applications in isolation, and that co-designing the software and hardware stack is critical. 

\vspace{-0.03in}
\subsection{Power Consumption}

Fig.~\ref{fig:battery_hivemind}a shows the consumed battery, on average, across drones by the end of execution; each of the 
10 jobs runs for 120s, and the two end-to-end scenarios run to completion; repeated 20 times. 
HiveMind consumes much less power compared to the distributed system, by offloading resource-demanding 
computation to the cloud. It also consumes less power than the centralized system, by avoiding 
excessive data transfer. Even though most power consumption is due to drone motion, communication can also exhaust 
the device's battery. There are two jobs ($S3$ and $S4$) for which HiveMind consumes slightly higher power than the centralized system. 
This is in line with our previous findings that these jobs do not benefit from dividing their execution between cloud and edge. 
Finally, the efficiency improvement for the end-to-end scenarios is largely due to HiveMind completing the scenario faster. 


\subsection{Network Bandwidth}

\begin{figure}
\centering
	\begin{tabular}{cc}
	\includegraphics[scale=0.168, viewport=70 64 810 400]{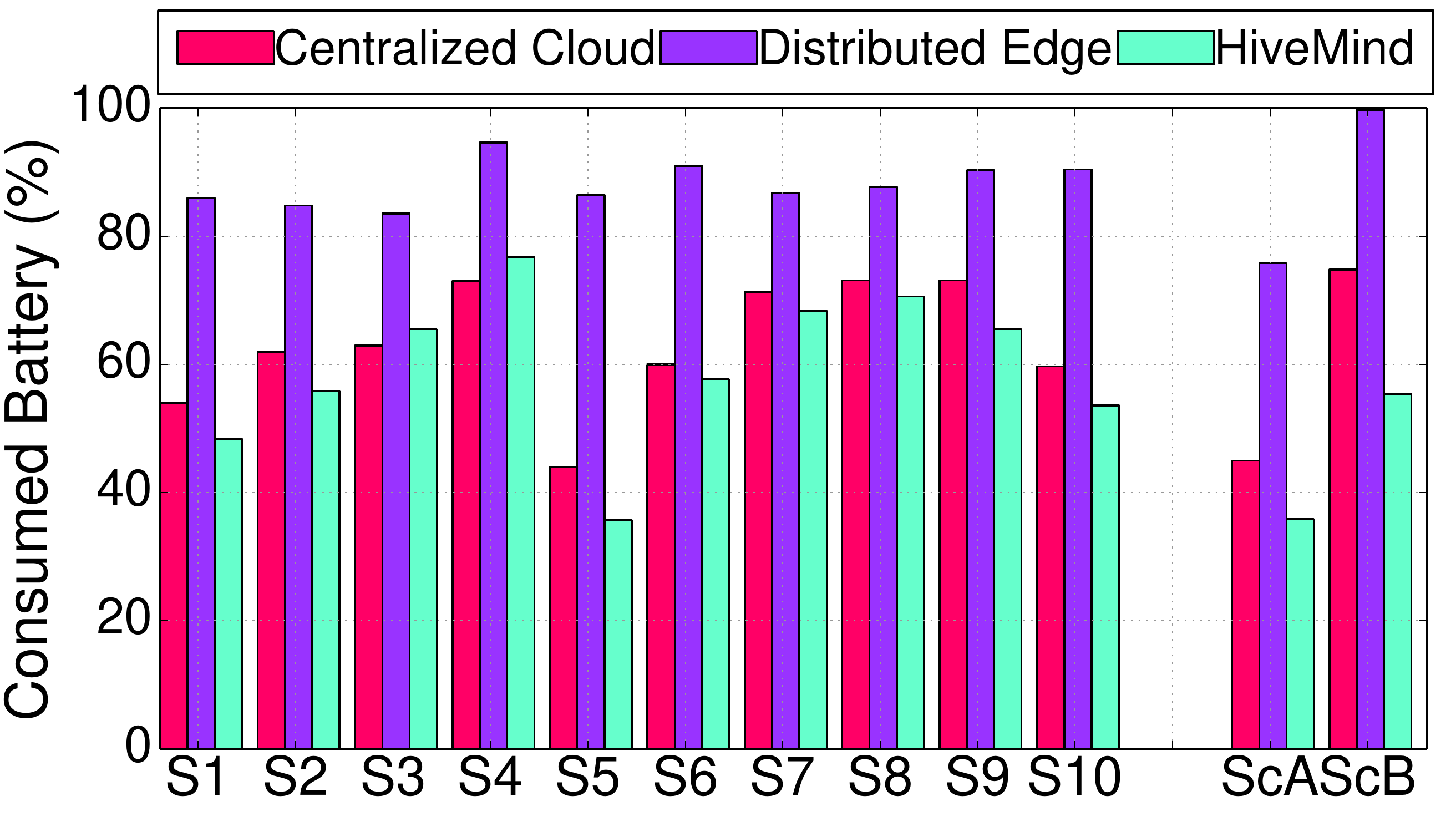} & 
	\includegraphics[scale=0.168, viewport=110 64 1010 400]{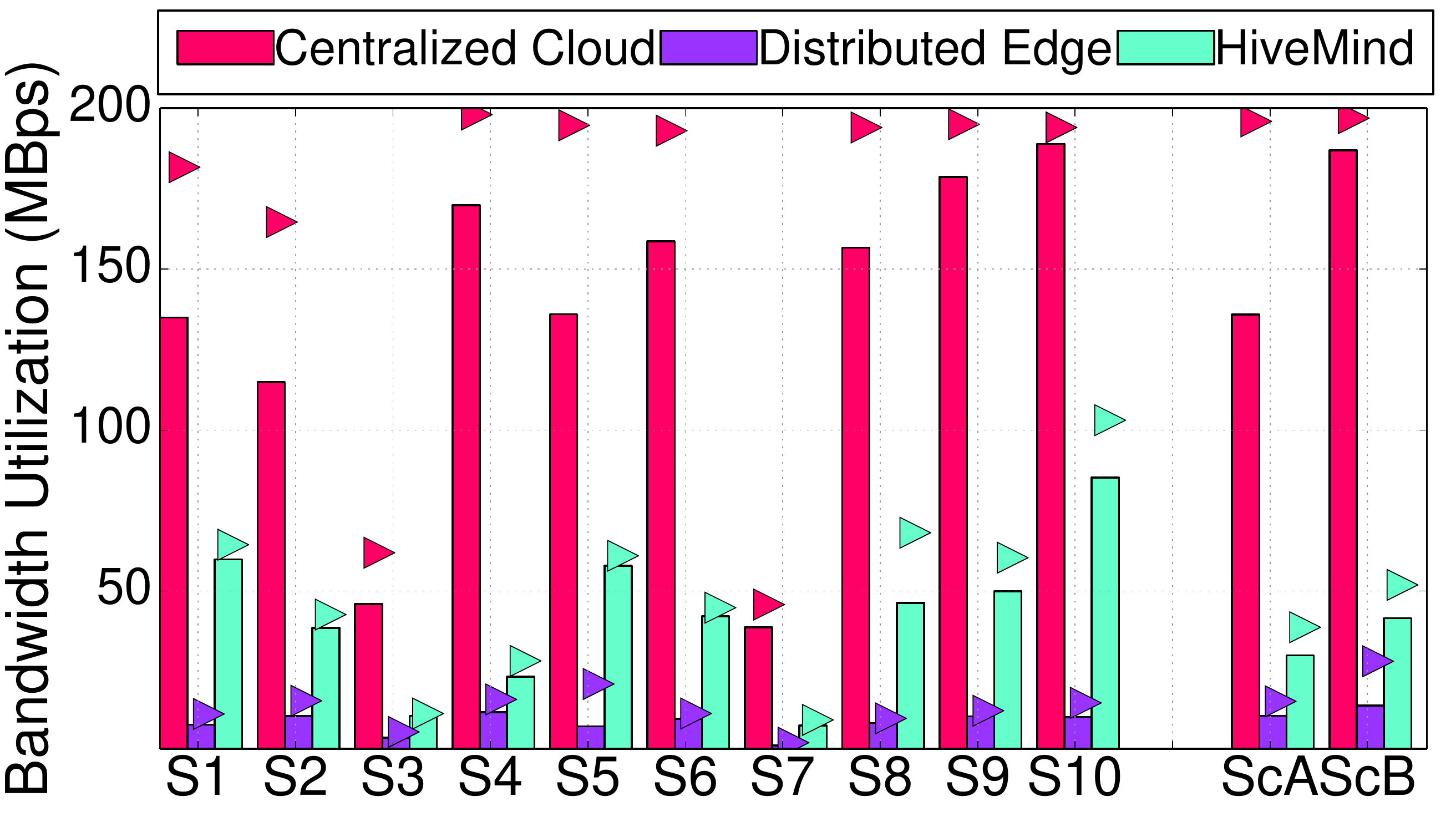} \\ 
	\end{tabular}
	\caption{\label{fig:battery_hivemind} Battery and network bandwidth consumption (bars$\rightarrow$median, markers$\rightarrow99^{th}$ \%ile) with HiveMind, the 
	centralized, and distributed platforms. }
	\vspace{-0.06in}
\end{figure}

Fig.~\ref{fig:battery_hivemind}b shows the network bandwidth usage across the three platforms. Bars show average and markers $99^{th}$ percentile bandwidth. 
While HiveMind consumes more bandwidth that the distributed system by offloading a fraction of data to the cloud, its usage 
is much lower than the centralized system, and it avoids the network congestion of offloading all data to the cloud. Note also that 
the difference between average and tail bandwidth is less pronounced for HiveMind, which contributes to its 
low performance variability. 


\begin{figure}
	\begin{tabular}{cc}
	\includegraphics[scale=0.20, viewport=50 54 700 438]{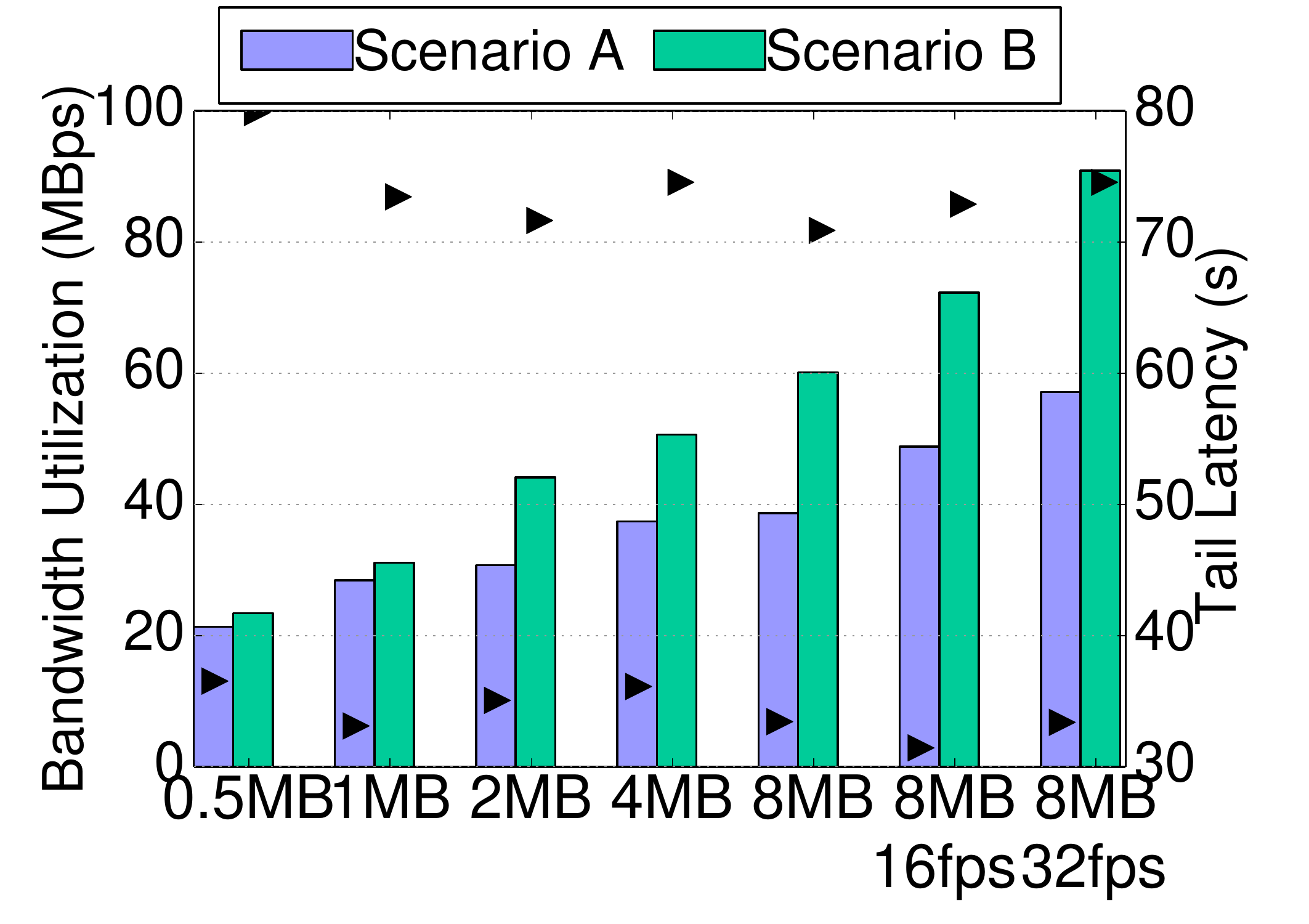} & 
	\includegraphics[scale=0.20, viewport=130 54 660 438]{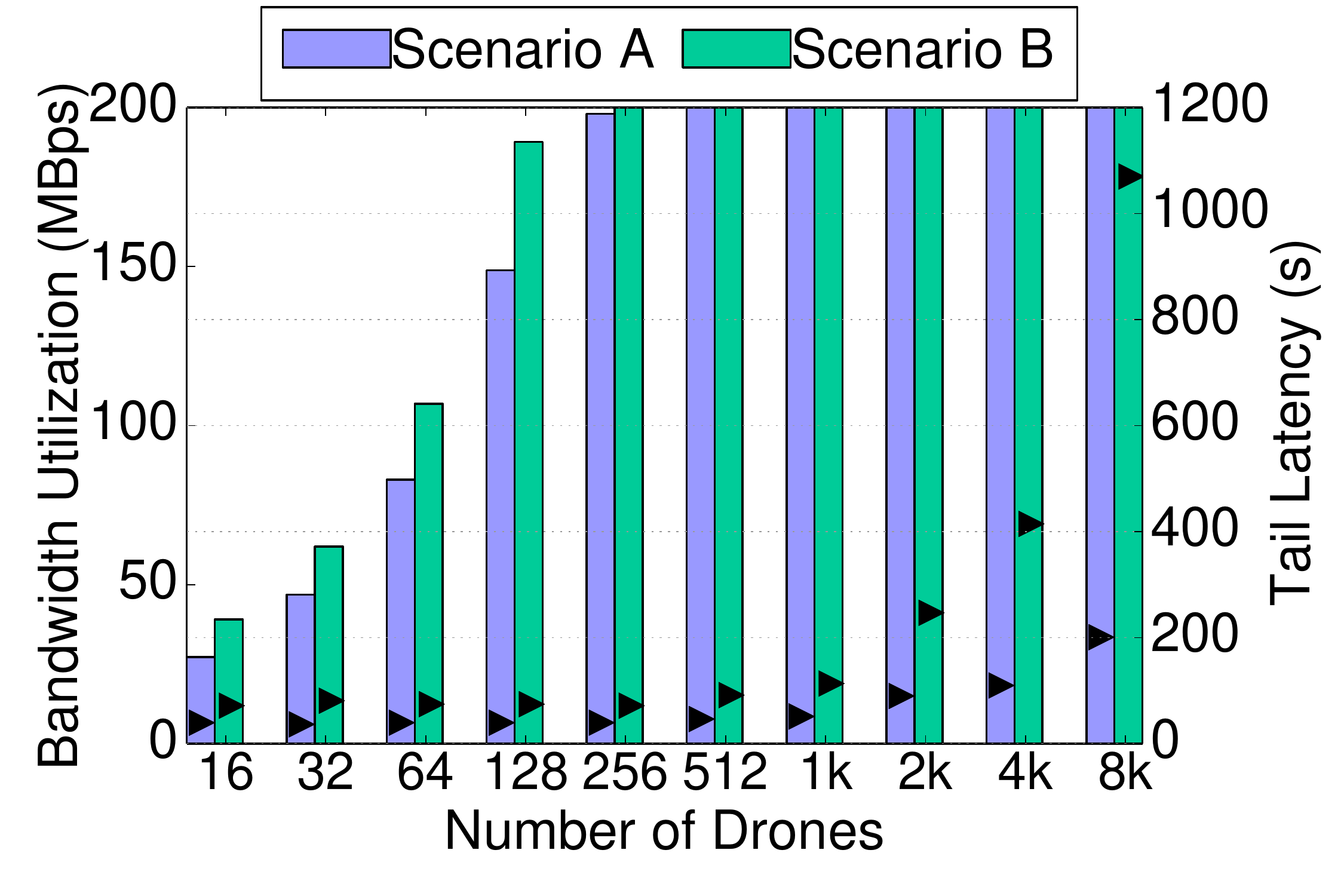} \\ 
	\end{tabular}
	\caption{\label{fig:scalability} Bandwidth (bars) and tail latency (markers) as we increase (a) resolution, and (b) \#drones. }
\end{figure}

\begin{figure}
\centering
	\begin{tabular}{ccc}
	\includegraphics[scale=0.16, viewport=60 44 660 370]{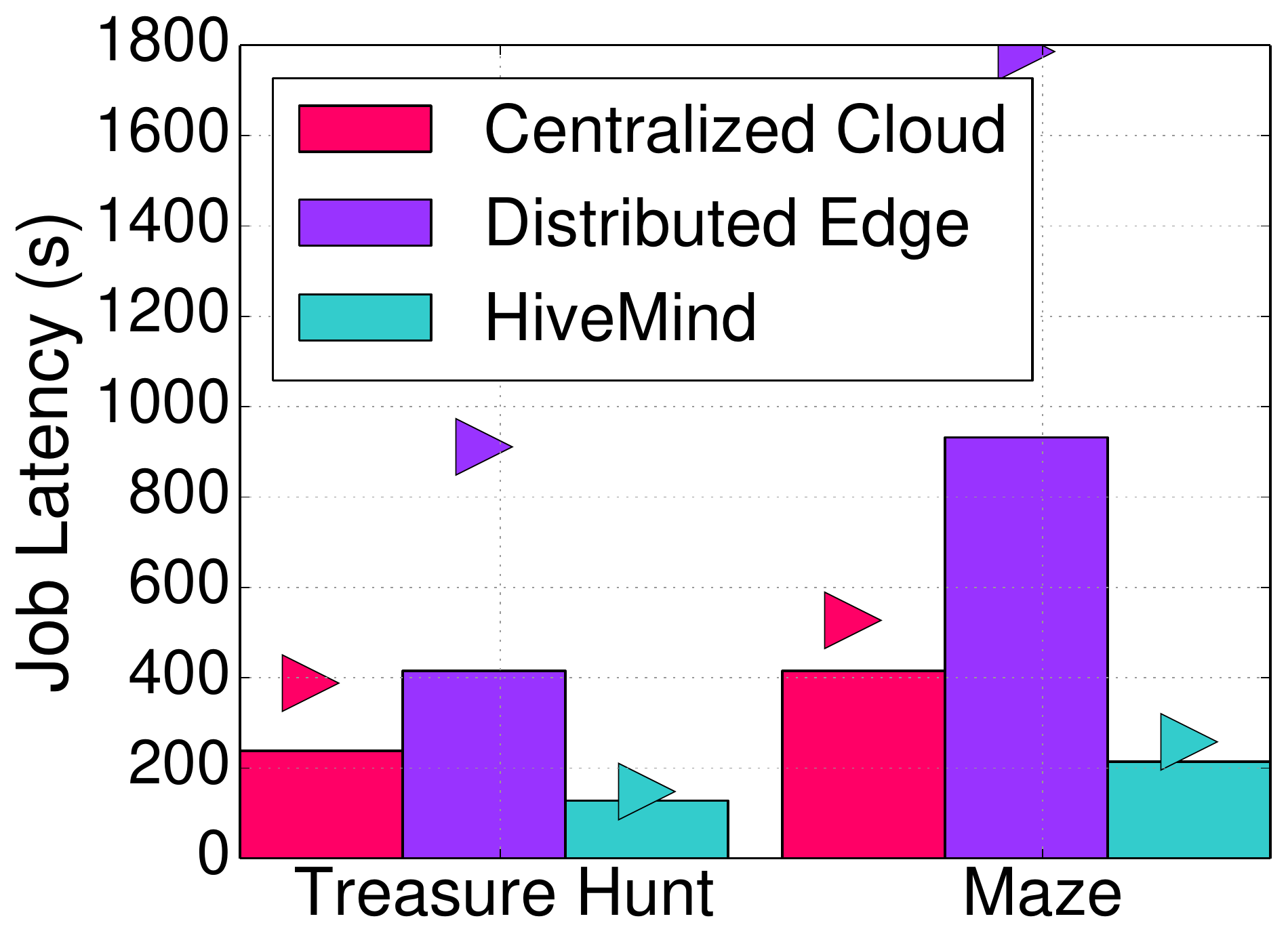} & 
		\includegraphics[scale=0.16, viewport=170 44 500 370]{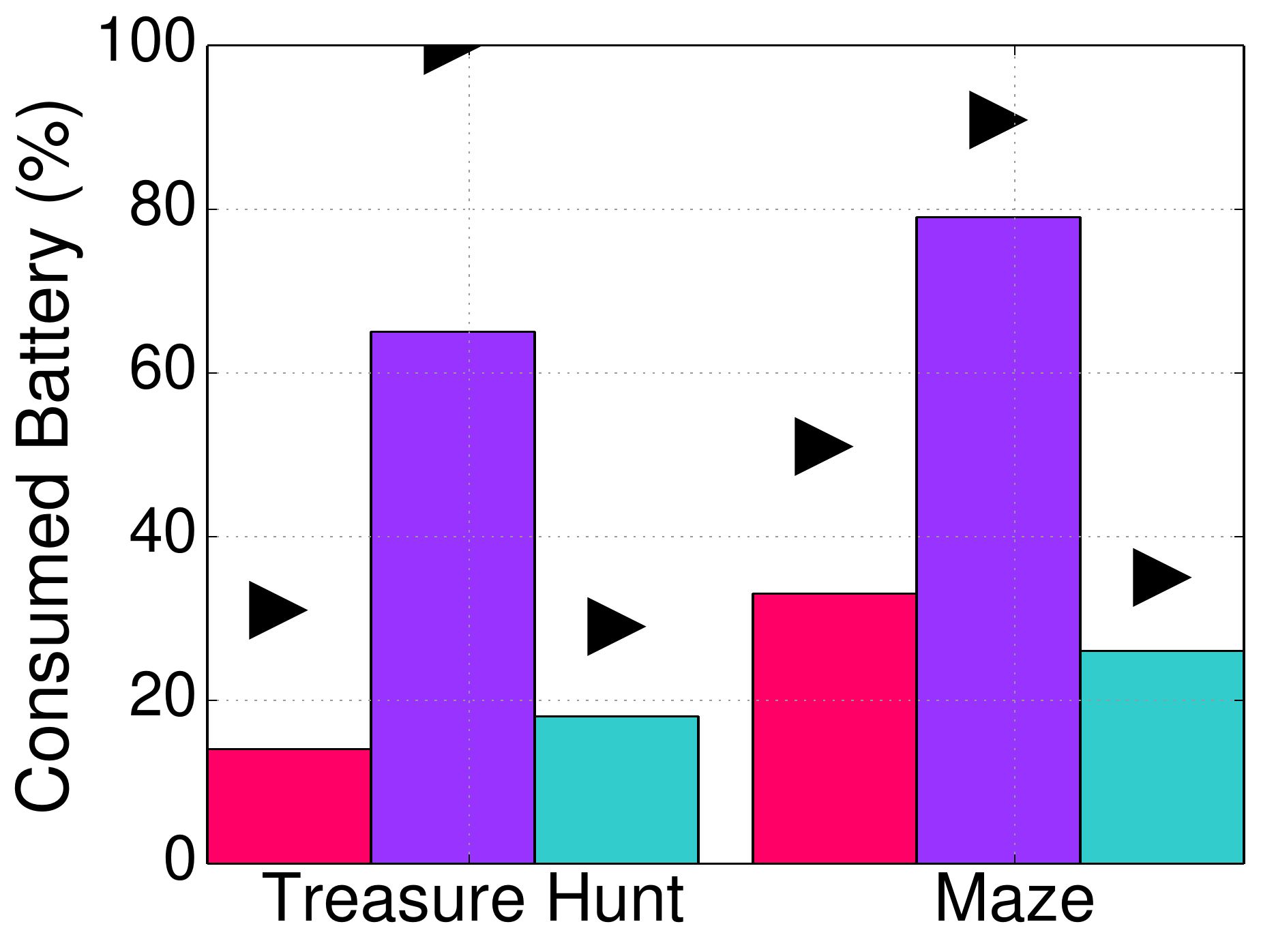} & 
        \includegraphics[scale=0.103, trim=13cm 0 0 0, clip=true]{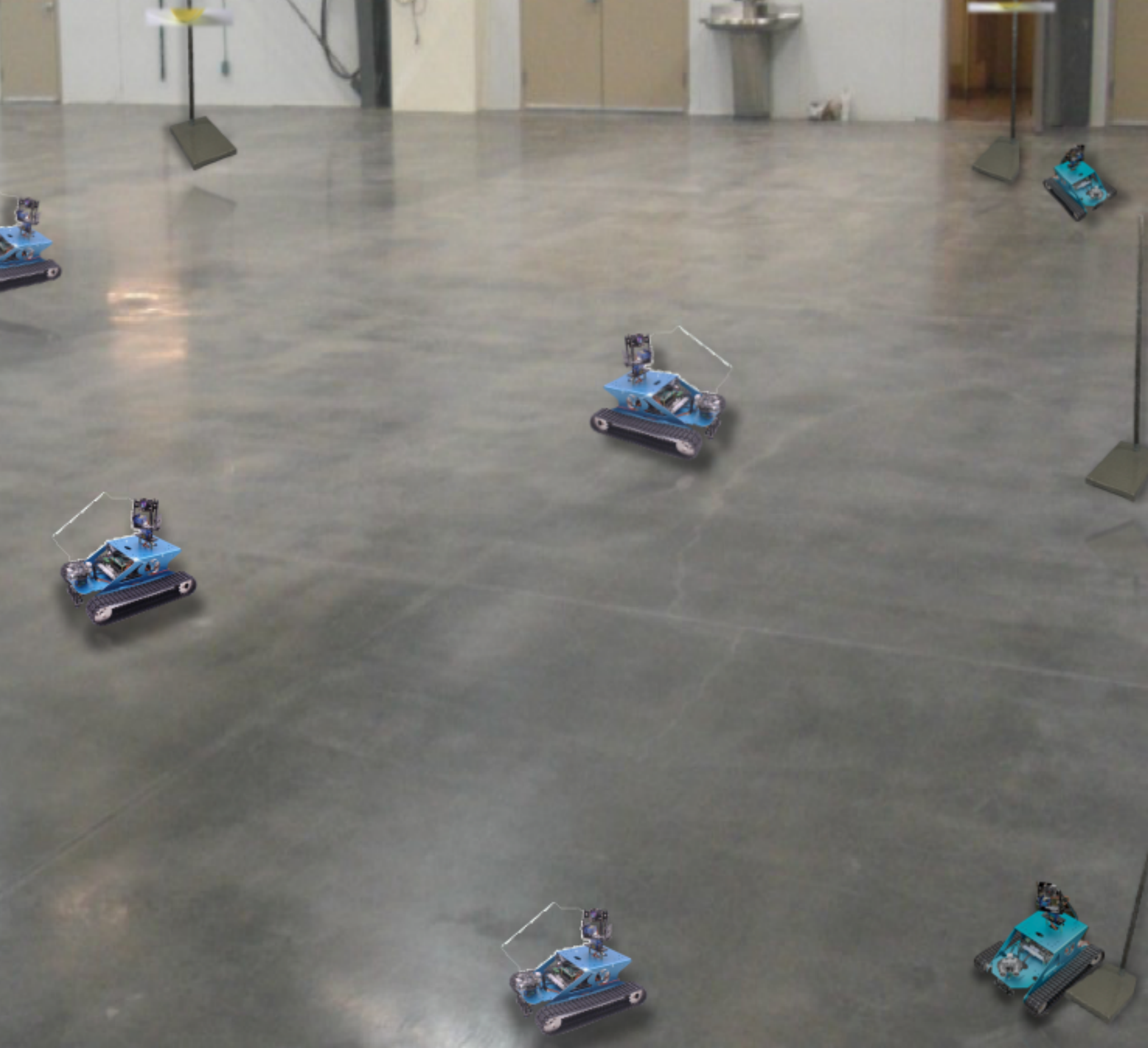}
	\end{tabular}
	\caption{\label{fig:robots} Latency and battery consumption of robotic cars (bars$\rightarrow$median, markers$\rightarrow$tail latency or battery consumption), 
	and subset of car swarm in ``Treasure hunt'' scenario. } 
	\vspace{-0.12in}
\end{figure}

\subsection{Applicability to Other Swarms}

HiveMind's design is not specific to drones. 
We now port HiveMind to a swarm of 14 robotic cars (Fig.~\ref{fig:robots}), each equipped with a high definition front camera, a Raspberry Pi board, 
and GPS, accelerometer, temperature, and altitude sensors~\cite{robot_car}. The serverless cluster is the same 
as before. Communication happens over a wireless network using TCP/IP. We explore two scenarios; 
a ``Treasure Hunt'', where, robots navigate a space with panels providing them instructions on where to move next until 
they reach a final target, and a ``Maze'', where they have to navigate an unknown maze. 
The first scenario involves image-to-text conversion to interpret the provided instructions. 
The user again expresses each scenario's task graph in HiveMind's DSL and provides 
the necessary task logic, and the system determines how to place tasks. The cars are less power-constrained 
than the drones, so obstacle avoidance and sensor analytics almost always run on-board. 

Fig.~\ref{fig:robots}a shows the median (bars) and tail (markers) job latency on HiveMind, and the centralized and distributed systems. 
Fig.~\ref{fig:robots}b shows the corresponding average (bars) and worst-case (markers) battery consumption. 
Performance is better and more predictable with HiveMind, especially compared to the distributed system. 
As with the drone swarm, the cars significantly benefit from network acceleration (22\% lower latency on average), but they also 
benefit from offloading expensive computation to the serverless cluster, without high instantiation overheads. 
Finally, because both scenarios have multiple phases, HiveMind's fast remote memory access 
significantly reduces their latency (19\% on average). 

\subsection{Scalability}
\label{sec:scalability}

Fig.~\ref{fig:scalability}a shows 
the bandwidth usage and tail latency for the two scenarios on the drone swarm, as the image resolution increases. 
Even for the maximum resolution and frame rate (32 fps), HiveMind does not saturate the network links, keeping latency low. 
In contrast, in the centralized system, network quickly became congested, 
only supporting modest resolutions and low frame rates (Fig.~\ref{fig:implications_net}). 

\begin{wrapfigure}[11]{r}{0.28\textwidth}
	\includegraphics[scale=0.19, viewport=60 44 660 370]{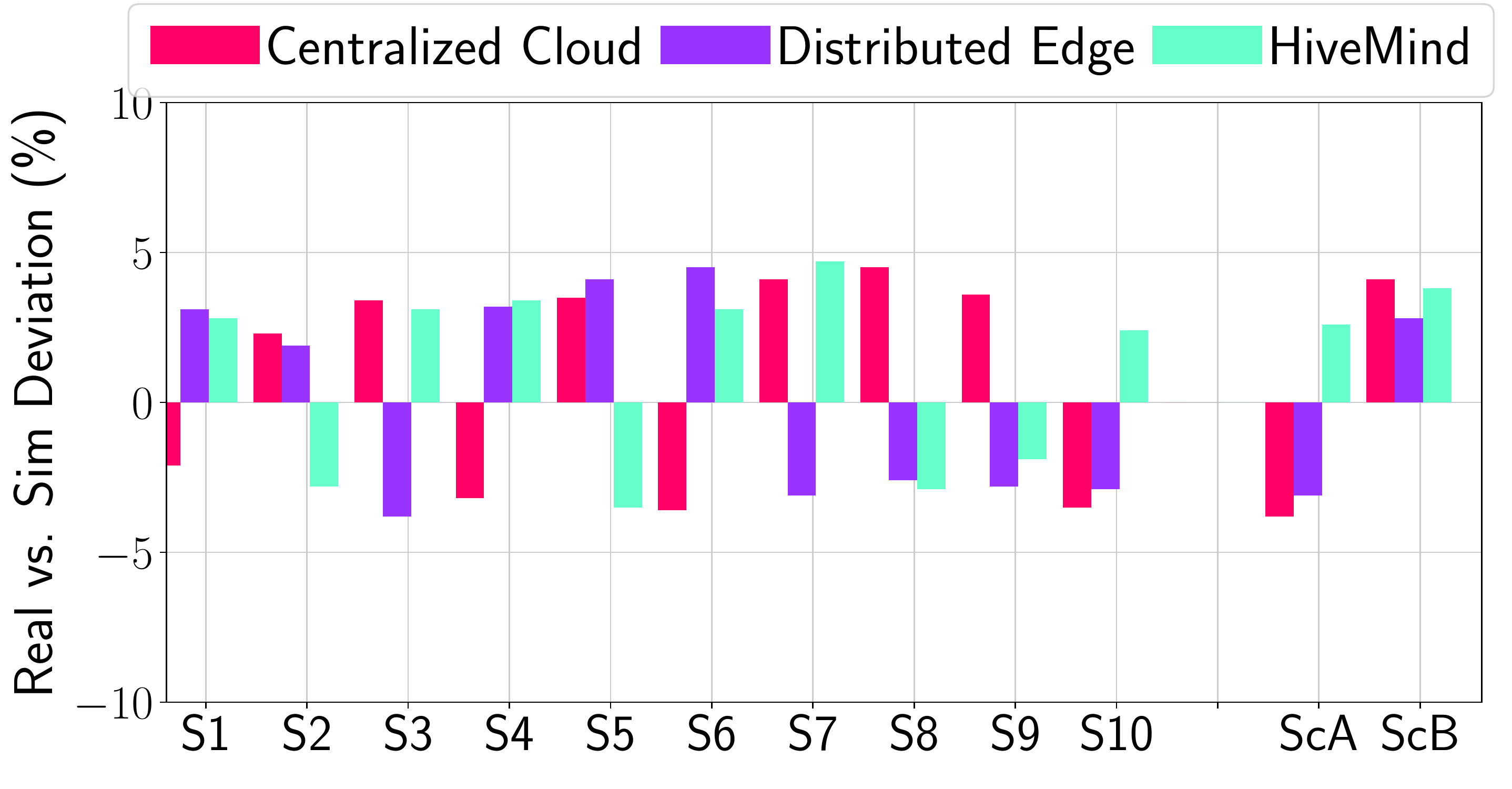}
	\caption{\label{fig:sim_validation} Validation of the simulator in terms of tail latency for the 16-drone swarm across the three configurations and all applications. }
\end{wrapfigure}
Since experimenting with hundreds of drones is impractical, we have also built a validated, event-driven simulator that 
accurately captures the performance, battery consumption, and network bandwidth of the real swarm. The simulator is based on 
queueing network principles and tracks the processing and queueing time both on cloud and edge resources. 
We have used the real drone and robotic car testbeds to validate the simulator's accuracy. 
Fig.~\ref{fig:sim_validation} 
shows a snapshot of the simulator's validation for the 16-drone swarm in HiveMind, the centralized, and distributed systems. We show the deviation 
in tail latency between the real experiments and simulation. In all cases deviation is less than 5\%. The results are similar for the robotic cars. 

We use the simulator to explore HiveMind's scalability to the swarm size. Fig.~\ref{fig:scalability}b shows the network bandwidth usage 
and tail latency for the two scenarios. We scale up the network links 
proportionately to the real experiments. Although for larger swarms the bandwidth usage increases, the increase 
is much slower than increase rate in devices, compared to a linear increase with the centralized system, especially 
for the first scenario, which accommodates more computation on-board. 









\section{Conclusion}
\label{sec:Conclusions}

We have presented HiveMind, a hardware-software system stack for edge swarms, which bridges the gap between centralized and distributed coordination. HiveMind implements a DSL to improve programmability for these systems, automatically handles task mapping between cloud and edge resources, and proposes hardware acceleration fabrics for remote memory access and networking. On real swarms with 16 drones and 14 robotic cars, HiveMind significantly outperforms prior systems, reducing overheads and abstracting away complexity. 

\balance
%

\bibliographystyle{IEEEtranS}
\bibliography{references}


\end{document}